\documentclass[useAMS,usenatbib,usegraphicx]{mn2e}

\def\ltsima{$\; \buildrel < \over \sim \;$}
\def\simlt{\lower.5ex\hbox{\ltsima}}
\def\gtsima{$\; \buildrel > \over \sim \;$}
\def\simgt{\lower.5ex\hbox{\gtsima}}
\def\gsimeq
{\hbox{\raise0.5ex\hbox{$>\lower1.06ex\hbox{$\kern-1.07em{\sim}$}$}}}
\def\lsimeq
{\hbox{\raise0.5ex\hbox{$<\lower1.06ex\hbox{$\kern-1.07em{\sim}$}$}}}
\def\pn{\par\noindent}
\def\ss{\smallskip\pn}

\def\xmm{{\it XMM-Newton }}

\def\xmm{{\it XMM-Newton}}

\def\ngc{NGC 4051}

\def\dchisq{\hbox{$\Delta\chi^2$}}

\title[X--ray spectral variability of NGC~4051] 
 {{\it XMM--Newton} study of the complex and variable spectrum of NGC~4051}

 \author[G.\ Ponti et al. ]
 {G.~Ponti$^{1,2,3}$\thanks{ponti@iasfbo.inaf.it},
   G. Miniutti$^{3}$, M. Cappi$^{2}$, L. Maraschi$^{4}$, A.C. Fabian$^{3}$ and K. Iwasawa$^{5}$ \\ \\
   $^1$Dipartimento di Astronomia, Universit\`a di Bologna, Via
   Ranzani 1, I--40127, Bologna, Italy\\
   $^2$INAF--IASF Sezione di Bologna, Via Gobetti 101, I--40129,
   Bologna, Italy \\
   $^3$Institute of Astronomy, Madingley Road, Cambridge CB3 0HA\\
   $^4$INAF, Osservatorio Astronomico di Brera, via Brera 28, 20121
   Milan,
   Italy\\
   $^5$Max--Planck--Institut f$\ddot{u}$r extraterrestrische Physik,
   Giessenbachstra$\ss$e, 85748 Garching, Germany}

\pagerange{\pageref{firstpage}--\pageref{lastpage}}
\pubyear{2001}

\usepackage{times}

\begin{document}

\label{firstpage}

 \maketitle

\begin{abstract}
  We study the X--ray spectral variability of the Narrow Line
  Seyfert~1 galaxy NGC~4051 as observed during two {\it{XMM--Newton}}
  observations. To gain insight on the general behaviour, we first
  apply model--independent techniques such as RMS spectra and
  flux-flux plots.  We then perform time--resolved spectral analysis
  by splitting the observations into 68 spectra (2~ks each).  The data
  show evidence for a neutral and constant reflection component and
  for constant emission from photoionized gas, which are included in
  all spectral models. The nuclear emission can be modelled both in
  terms of a ``standard model'' (pivoting power law plus a black body
  component for the soft excess) and of a two--component one (power
  law plus ionized reflection from the accretion disc).  Both models
  reproduce the source spectral variability and cannot be
  distinguished on a statistical ground. The distinction has thus to
  be made on a physical basis. The standard model results indicate
  that the soft excess does not follow the standard black body law
  (L$_{BB} \propto T^4$) despite a variation in luminosity by about
  one order of magnitude.  The resulting temperature is consistent
  with being constant and has the same value as observed in PG
  quasars.  Moreover, although the spectral slope is correlated with
  flux, which is consistent with spectral pivoting, the hardest photon
  indexes are so flat ($\Gamma\sim$1.3--1.4) as to require rather
  unusual scenarios. Furthermore, the very low flux states exhibit an
  inverted $\Gamma$--flux behaviour which disagrees with a simple
  pivoting interpretation.  These problems can be solved in terms of
  the two--component model in which the soft excess is not thermal,
  but due to the ionized reflection component. In this context, the
  power law has constant slope (about 2.2) and the slope--flux
  correlation is explained in terms of the relative contribution of
  the power law and reflection components which also explains the
  shape of the flux--flux plot relationship. The variability of the
  reflection component from the inner disc closely follows the
  predictions of the light bending model, suggesting that most of the
  primary nuclear emission is produced in the very innermost regions,
  only a few gravitational radii from the central black hole.
\end{abstract}

\begin{keywords}
  galaxies: individual: NGC~4051 -- galaxies: active -- galaxies:
  Seyfert -- X-rays: galaxies
\end{keywords}

\section{Introduction}

NGC~4051 is a nearby (z=0.0023) low--luminosity Narrow--Line Seyfert 1
(NLS1) galaxy which exhibits extreme X--ray variability in flux and
spectral shape on both long and short timescales. The source sometimes
enters relatively long and unusual low flux states in which the hard
spectrum becomes very flat ($\Gamma \sim 1$) while the soft band is
dominated by a much steeper component ($\Gamma \sim 3$, or blackbody
with \ $kT\sim 0.12$~keV). Most remarkable is the 1998 {\it BeppoSAX}
observation reported by Guainazzi et al (1998) in which the source
reached its minimum historical flux state, the spectral slope was
$\Gamma \simeq 0.8$, and the overall spectrum was best explained by
assuming that the nuclear emission had switched off leaving only a
reflection component from distant material (clearly visible in the
hard spectrum) which still responded to previous higher flux levels. A
different interpretation assumes that the nuclear emission in low flux
states originates so close to the central black hole that only a tiny
fraction can escape the gravitational pull of the hole so that the
nuclear continuum is virtually undetectable at infinity (e.g.
Miniutti \& Fabian 2004).

In general, the 2--10~keV spectral slope appears to be well correlated
with flux. However, the correlation is not linear with the slope
hardening rapidly at low fluxes and reaching an asymptotic value at
high fluxes (see e.g. Lamer et al 2003, based on {\it RXTE} monitoring
data over about three years).  This behaviour is relatively common in
other sources as well, such as MCG--6-30-15, 1H~0419--577, NGC~3516
(among many others) and might be due to flux--correlated variations of
the power law slope produced in a corona above an accretion disc as
originally proposed by Haardt \& Maraschi (1991; 1993) and Haardt et
al. (1994) and related to changes in the input soft seed photons (e.g.
Maraschi \& Haardt 1997, Haardt, Maraschi \& Ghisellini 1997, Poutanen
\& Fabian 1999, Zdziarski et al 2003). On the other hand, such
slope--flux behaviour can be explained in terms of a two--component
model (McHardy et al 1998; Shih, Iwasawa \& Fabian 2002) in which a
constant slope power law varies in normalisation only, while a harder
component remains approximately constant hardening the spectral slope
at low flux levels only, when it becomes prominent in the hard band.

In the well studied case of MCG--6-30-15, detailed spectral and
variability analysis (e.g. Fabian \& Vaughan 2003; Vaughan \& Fabian
2004) allowed the presence of two main spectral components to be
disentangled, namely a variable power law component (PLC) with
constant slope, and a reflection--dominated component (RDC) from the
inner accretion disc which does not follow the PLC variation,
confirming for this particular case the validity of the two--component
model and ruling out spectral pivoting as the main source of spectral
variability. The contribution from a distant reflector is limited to
10 per cent or less (see e.g. the long {\it Chandra} observation
reported by Young et al 2005) i.e. the RDC is completely dominated by
lowly ionized reflection from the inner disc, including the broad
relativistic Fe line.  From a theoretical point of view, the lack of
response of the disc RDC to the PLC variations has been interpreted in
terms of strong gravitational light bending by Miniutti et al (2003)
and Miniutti \& Fabian (2004), which predicts this behaviour
during normal flux periods and gives a correlation at low flux
levels only. 

It should be stressed that, in the framework of the two--component
model, the 2--10~keV slope--flux correlation in NGC~4051 cannot
entirely be explained by the contribution of a distant reflector in
the hard band (the presence of which is required by the detection of a
narrow component to the iron line). Indeed, as shown by Lamer et al
(2003) the correlation is still present even if the distant reflector
contribution is taken into account. Even at low flux levels,
soft and hard components are significantly variable and well
correlated, which excludes any dominant extended emission
(torus and/or extended scattering region). Extended emission is 
undetected by {\it Chandra} on $\sim$~100~pc scales (Collinge et al
2001; Uttley et al 2003). Therefore, the residual slope--flux
correlation can be due to intrinsic pivoting or to an additional,
almost constant, spectral component (i.e. RDC from the disc, as for
MCG--6-30-15). Throughout the paper for ``two--component model'' we
mean a parametrization of the nuclear emission in terms of two
components. Other spectral components such as absorption, emission
from photoionized gas, and reflection from distant matter can be
present but do not represent nuclear emission.

Other important clues on the nature of the spectral variability in
NGC~4051 (and other sources as well) comes from the so--called
flux--flux plot analysis, first presented by Churazov, Gilfanov \&
Revnivtsev (2001) to demonstrate the stability of the disc emission in
the high/soft state of Cyg--X1. Taylor, Uttley \& McHardy (2003) have
applied the same technique to a few Seyfert galaxies by using {\it
  RXTE} data.  In the case of NGC~4051, Uttley et al (2004) have
performed such an analysis by using the same {\it XMM--Newton} data we
are presenting here and showed that the distribution of high,
intermediate, and low flux state data points are smoothly joined
together, indicating that the same process causing the spectral
variability in high and intermediate flux states continues to operate
even at the very low flux levels probed by the {\it XMM--Newton} data
(Uttley et al 2004). This evidence seems to challenge the
interpretation of the spectral variability as due to variable
absorption by a substantial column of photoionized gas (Pounds et al.
2004).

Here we report results based on the two {\it XMM--Newton} observations
of NGC~4051. The data have been previously analysed by Uttley et al
(2004) and Pounds et al (2004) who reach different conclusions on the
nature of the spectral variability in NGC~4051 and on the main
spectral components that come into play. Our analysis is complementary
to previous studies and offers a novel point of view that may also be
relevant for other similar sources.

\begin{figure*}
\hbox{
 \includegraphics[width=0.34\textwidth,height=0.46\textwidth,angle=-90]{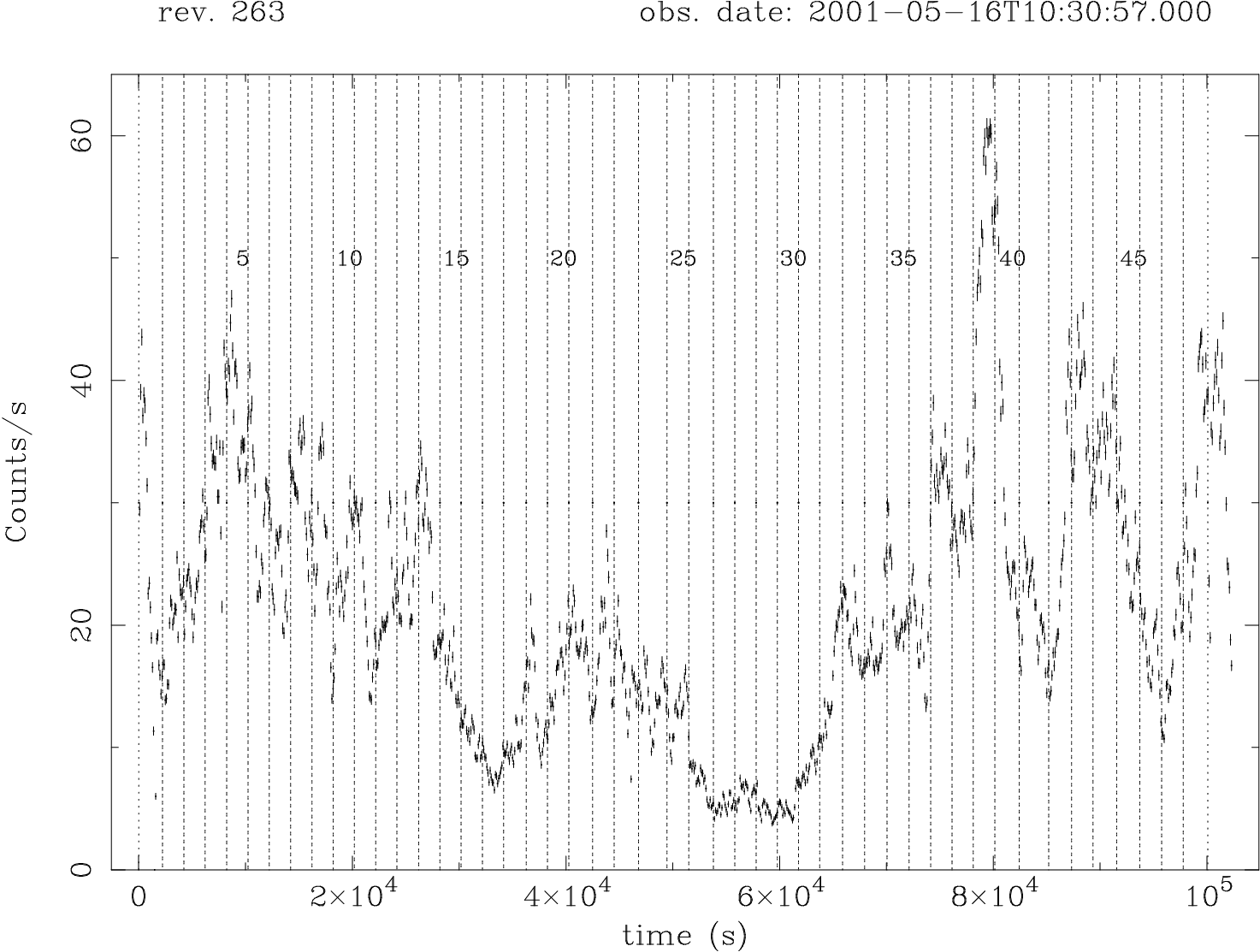}
\hspace{1.0cm}
 \includegraphics[width=0.34\textwidth,height=0.46\textwidth,angle=-90]{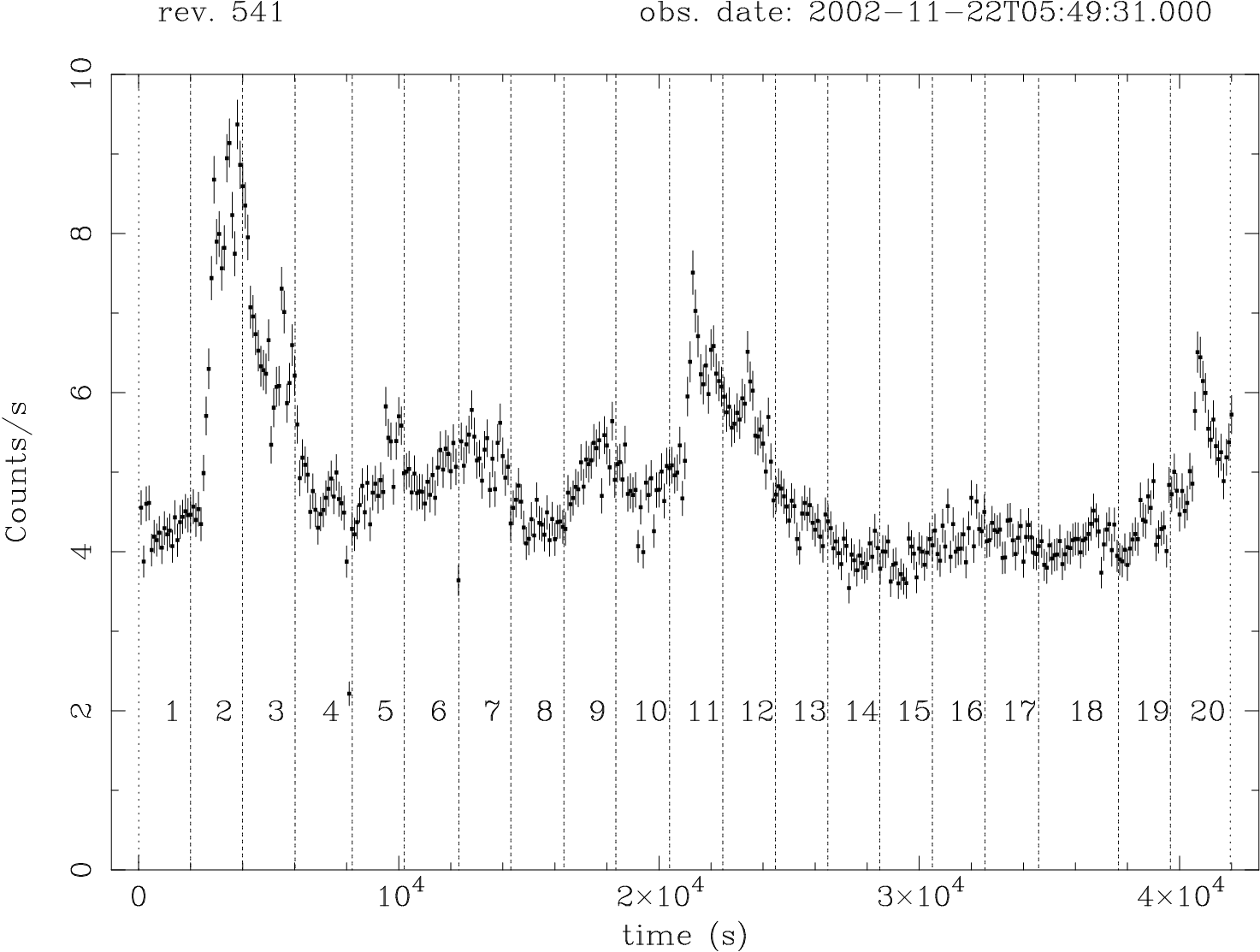} 
}
\caption{The broadband 0.2--10~keV {\it XMM--Newton} EPIC--pn light
  curves of NGC~4051 for the first (left panel, rev.~263) and the
  second (right panel, rev.~541) observations. The 2~ks time slices
  used in the time--resolved spectroscopic analysis are also
  shown.}
\label{lcs}
\end{figure*}

\section{The two XMM--Newton observations}

NGC~4051 has been observed twice by {\it XMM--Newton}. The first
observation took place during revolution 263 (2001 May) for $\sim$117
ks, the second during revolution 541 (2002 November) for $\sim$52 ks.
These data are public and have been obtained from the \xmm~ data
archive.  The analysis has been made with the latest version of the
SAS software, starting from the ODF files.  During the first
observation the EPIC pn and the MOS2 were operated in small window
mode with the medium filter.  No significant pile-up is present in
the pn data, so single and double pixel events were selected.  The
MOS2 data show $\sim$1\% of pile-up when the count rate is higher than
5 counts/s (during the first observation it sometimes reaches 15
counts/s).  This could invalidate the results of time--resolved
spectral variability, because, even if pile-up is not strong, its
effect is increasing with flux.  Thus, for the time--resolved spectral
analysis, we use the MOS2 data selecting only single events and
extracting the source photons within a circular region of
30$^{\prime\prime}$ radius when the count rate is lower than 5
counts/s, and within an annular region with 7$^{\prime\prime}$ and
30$^{\prime\prime}$ radii, during the higher count rate intervals.
The MOS1 data were in Fast Uncompressed mode and, due to cross
calibration uncertainty, we did not used those.  The second
observation was timed to coincide with an extended period of low
X--ray emission from \ngc.  During this observation the pile-up is
negligible in all the EPIC cameras and single plus double pixel events
are used for MOS1, MOS2 and pn cameras.

The background spectrum has been taken from source--free regions in
the same chip as the source and ancillary and response files have been
created. After filtering out periods of high background, we obtain net
exposures of about 97~ks ($\sim$69~ks for the pn small window mode)
and 39~ks for the first and second observation respectively. At the
moment the pn camera is not well calibrated below 0.5~keV.  For the
data we are presenting here, for example, the ratio between the MOS
and pn camera below 0.5~keV is of about 20 per cent.  Thus, the data
from the pn camera are used in the 0.5--10~keV band only, while MOS
data are used down to 0.2~keV. All spectral fits include absorption
due to the \ngc~ line-of-sight Galactic column of
$N_H$=1.32$\times$10$^{20}$ cm$^{-2}$ (Elvis et al. 1989).  In the
subsequent spectral analysis, errors are quoted at the 90 per cent
confidence level (\dchisq=2.7 for one interesting parameter).

\section{Spectral variability: a quick look}

The {\it XMM--Newton} broadband light curve of NGC~4051 exhibits large
amplitude count rate variations in both observations, typical
for this source and for NLS1 galaxies in general. Fast large amplitude
variability is superimposed on longer trends which are mainly
characterized by persistent low flux periods in which variability
is suppressed. The {\it XMM--Newton} light curves shown in
Fig.~\ref{lcs} are a good representation of this behaviour which can
be described in more general terms by the RMS--flux relation in which
the absolute RMS variability is proportional to the source flux
(Uttley \& M$^{\rm c}$Hardy 2001). Light curves such as those shown in
Fig.~\ref{lcs} are just a realization of the RMS--flux relation which
naturally produces large amplitude spikes at high fluxes and periods
of relative quiescence at low fluxes (e.g. Uttley, M$^{\rm c}$Hardy \&
Vaughan 2005).

The general shape of the broadband spectrum of NGC~4051 in both
observations can be roughly described by the presence of a power law
continuum, reflection from distant matter including a narrow 6.4~keV
Fe line, and a prominent soft excess below about 1~keV. The Fe line is
unresolved and fluxes are consistent with each other in the two
observations.  A Warm Absorber imprints its presence in the spectrum
particularly through two edge like features at 0.74 and 0.87~keV
(O~{\footnotesize{VII}} and O~{\footnotesize{VIII}}).  Even modelling
the absorption, some structures persist between 0.8~keV and 1~keV,
particularly evident in the low flux state.  This further structure
can be interpreted as a broad absorption structure (as suggested by
Pounds et al 2004), emission around 0.9~keV (see Collinge et al 2001,
Uttley et al 2004), or a combination of the two.

\begin{figure}
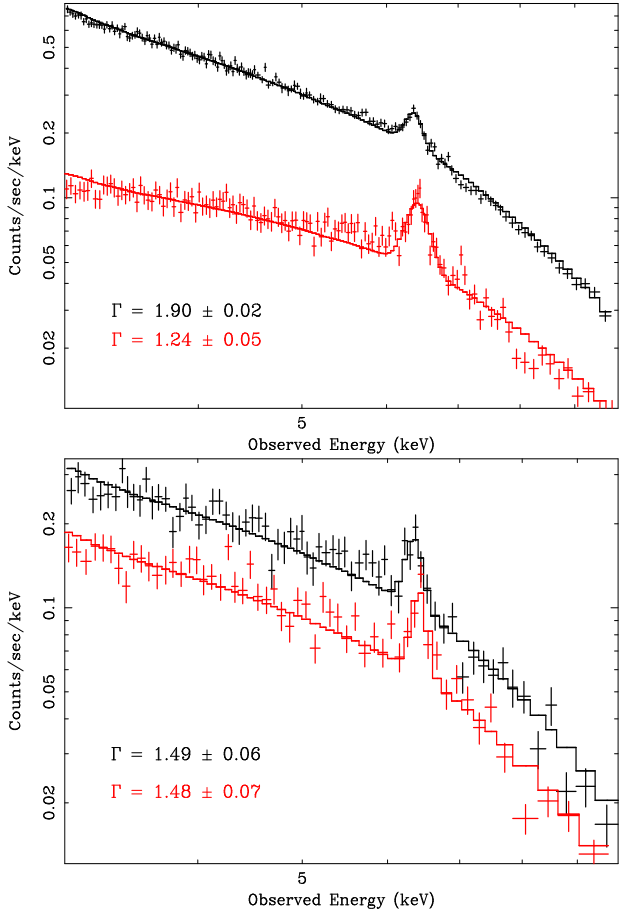

 \includegraphics[width=0.34\textwidth,height=0.46\textwidth,angle=-90]{medio.cps}
 \includegraphics[width=0.34\textwidth,height=0.46\textwidth,angle=-90]{basso.cps} 
 \caption{In the top panel we show the time averaged 3--10~keV for the
   first and second observation. The spectrum in the second low flux
   observation is flatter and much more curved that during the first
   high flux one. In the bottom panel we show two $\sim$~10~ks spectra
   extracted from the first and second observation respectively. The
   one from the first observation is taken around 60~ks (see right
   panel of Fig.~\ref{lcs}) and is representative of the low flux
   state during that (higher mean flux) observation. The two spectra
   are chosen from intervals that have approximately the same count
   rate. The two spectra have the same spectral slope and the same
   hard curvature showing that the spectral variability is clearly
   flux--dependent and that the same flat spectrum and curvature
   characteristic of the second observation can be found also in the
   first, if low flux levels are selected.}
\label{test}
\end{figure}

Spectral variability is clearly present between the two observations
with the source being much harder in the second, lower flux
observation.  Moreover, when the source is faint and hard, the
spectrum above $\sim$~2~keV appears to be more curved than in the high
state. To illustrate the differences in spectral slope and hard
curvature between the two observations we show the two 3--10~keV
time--averaged spectra in the top panel of Fig.~\ref{test}. The
best--fitting power law slope is indicated, the fit including a narrow
6.4~keV Fe line as well. The spectral curvature can be interpreted as
the effect of absorption by a relatively high column of ionized gas or
could be the signature of a relativistically blurred reflection
component from the accretion disc.

Two main mechanisms have been proposed so far to explain the spectral
variability of NGC~4051. Pounds et al (2004) performed a comparative
analysis of the two {\it XMM--Newton} observations based on the
time--averaged spectra and on the simultaneous high--resolution RGS
data in the soft band.  Their main conclusions is that the spectral
variability between the high flux (rev.~263) and low flux (rev.~541)
observations is mainly due to variable absorption by a substantial
column of photoionized gas in the line of sight. The authors suggest
that in the low state (observed about 20 days after the source entered
one of its characteristic low state periods) part of the gas had
recombined in response to the extended period of lower X--ray flux so
providing a partial coverer (with $N_H \sim 3.6\times
10^{23}$~cm$^{-2}$). The variable opacity of the absorber could then
explain in a natural way both the flat spectrum and the hard curvature
that is observed in the second low flux observation.

This interpretation is challenged by the analysis by Uttley et al
(2004) who proposed spectral pivoting as the main driver of the
spectral variability in NGC~4051 and pointed out that a variation in
the absorber opacity in the line of sight during the low state is not
consistent with the nature of the spectral variability, as revealed by
flux--flux plot analysis, in which no sharp transition is seen between
the two observations. The flux--flux relationship rather indicates
that the spectral variability is flux--dependent and the low flux
states of the first observation are virtually identical in spectral
shape to the high flux state of the second. This would rule out that
the low flux spectral shape during the second observation depends on
the fact that data were obtained following an extended low flux state
allowing the absorbing gas to recombine.

It is clear that the Pounds et al interpretation implies that the
spectral shape is different in the two observations. For instance, the
low flux states of the first (high flux) observation should not be as
flat and curved as the spectrum in the second observation. This idea
can be easily checked by comparing two spectra: the first one is
representative of the very low flux state during the first observation
(around 60~ks, see Fig.~\ref{lcs}) and the second one is extracted
from the second observation. We choose the two spectra to have the
same exposure and approximately the same count rate. The two spectra
are shown in the bottom panel of Fig.~\ref{test} in the hard band. The
best fitting power law slope is also indicated. The fit also includes
Fe emission at 6.4~keV. Firstly, the spectral slope is exactly the
same in the two spectra. Therefore the slope is count--rate dependent
and the time--averaged flat spectrum during the low flux observation
has thus little to do with the extended period of low flux preceding
the second observation.  Secondly, the two spectra also seem to have
approximately the same hard curvature. When fitted with a partial
covering model, both spectra require a large column density of $\sim
0.5$--$2 \times 10^{23}$~cm$^{-2}$ and a covering fraction of 60--70
per cent. This demonstrates that not only the apparent spectral slope
is flux--dependent, but also that the hard curvature depends on flux
only.

From the above analysis, it seems likely that the spectral variability
in NGC~4051 is simply flux--dependent and that the second low flux
observation is not peculiar but indeed identical to low flux states
during the first observation. As already mentioned by Uttley et al
(2004) one possible explanation for the flux--dependent spectral
variability is spectral pivoting. In this paper we shall also explore
an alternative explanation invoking a reflection component from the
accretion disc.  Before considering time--resolved spectral analysis,
we perform some model--independent tests with the aim of identifying
the main spectral components and their relative contribution to the
spectrum and variability, in order to have a guide for our subsequent
analysis.

\section{RMS spectra}

\begin{figure}
 \includegraphics[width=0.34\textwidth,height=0.46\textwidth,angle=-90]{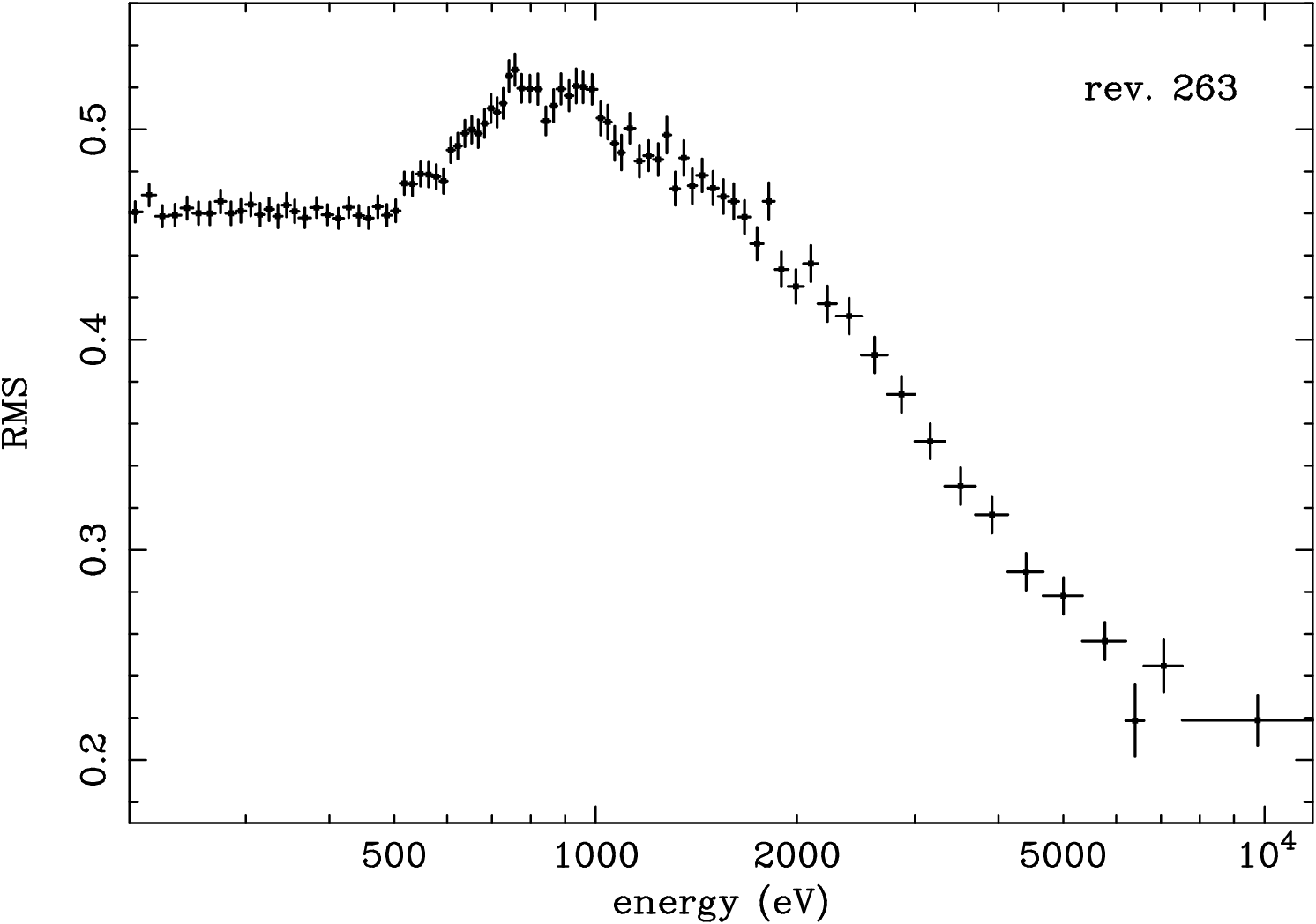}
 \includegraphics[width=0.34\textwidth,height=0.46\textwidth,angle=-90]{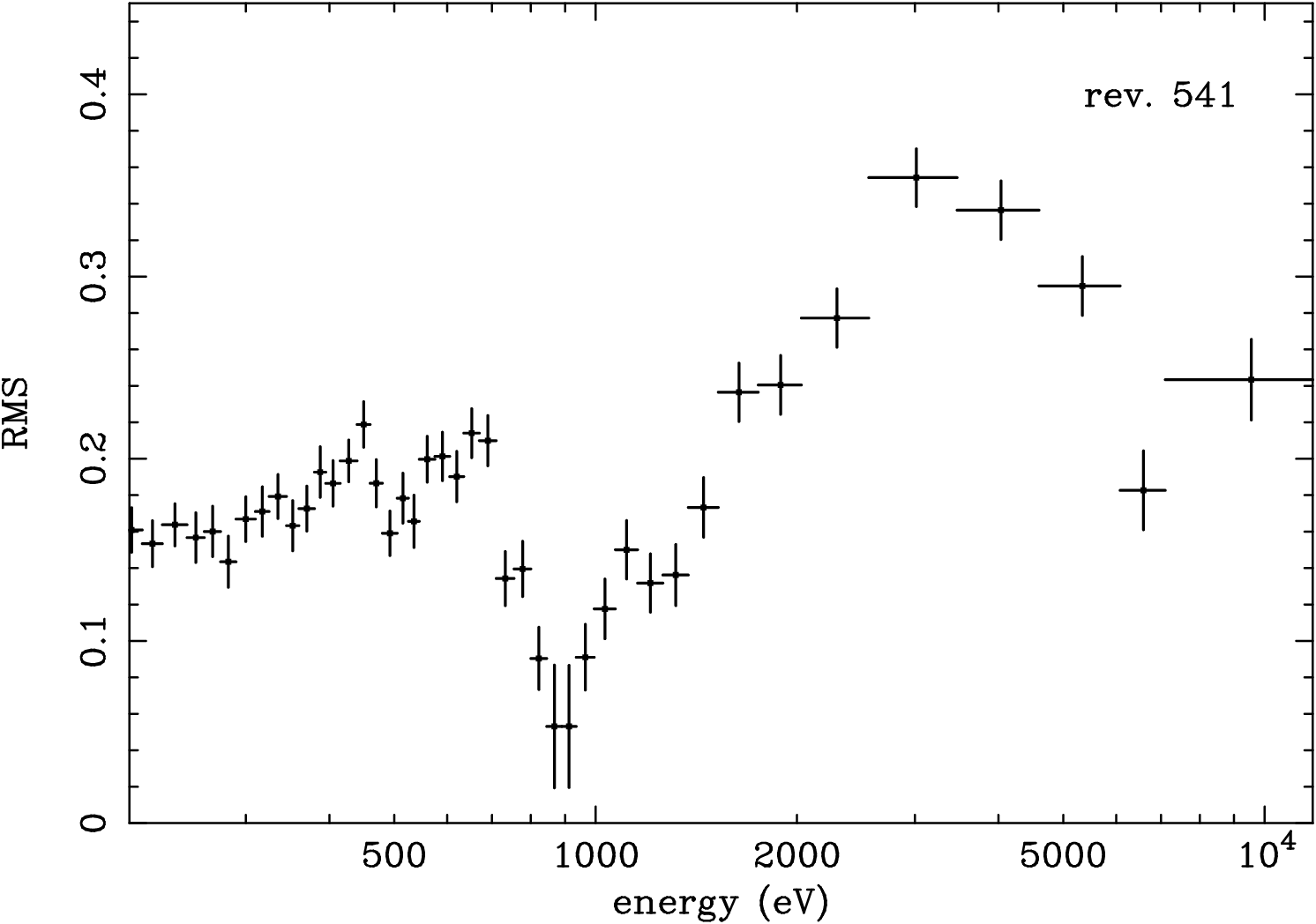} 
 \caption{The RMS spectra of the rev. 263 (top panel) and rev. 541
   (bottom panel). The RMS spectra are computed with time bins of 2~ks
 with a minimum of 300 counts per energy bin.}
\label{RMS}
\end{figure}

We start our analysis of the spectral variability of \ngc~ by
calculating the RMS spectrum, which measures the
total amount of variability as a function of energy (Edelson et al. 2002; 
Vaughan et al. 2004; Ponti et al. 2004).
Fig. \ref{RMS} shows the RMS spectra during  rev.~263 and 541.
The RMS spectra have been calculated with time bins of $\sim$2 ks (the
same as the subsequent time--resolved spectral analysis) and with energy bins
chosen in order to have at least 300 counts per energy--time bin. In
this way the poissonian noise is negligible compared to the RMS value and 
thus the use of the uncertainty given in Ponti et al. (2004) is valid.  

During the high flux observation of rev.~263 (Fig.~\ref{RMS}) the
variability rapidly increases toward softer energies, i.e.  the
broadband emission tends to steepen as it brightens.  Some reduction
in the variability is seen below about 800~eV followed by a plateau
below 500~eV. A similar RMS shape has been observed in other
sources (Ponti et al. 2005; Vaughan et al. 2003; Vaughan et al. 2004)
and the two simplest explanations for the broadband trend invoke
either spectral pivoting of the variable component, or the
two--component model (see e.g Markowitz, Edelson \& Vaughan 2003).
Another important feature of the RMS spectrum is a drop of variability
at 6.4~keV, the energy of the narrow Fe line which is seen in the
time--averaged spectrum. Such a drop shows that the narrow Fe line (and
therefore the associated reflection continuum) is much less variable
than the continuum (if variable at all) indicating an origin in
distant material. Some structure is also present around 0.9~keV, with
the possible presence of either a drop at that energy or two peaks.

\begin{figure}
\begin{center}
 \includegraphics[width=0.34\textwidth,height=0.46\textwidth,angle=-90]{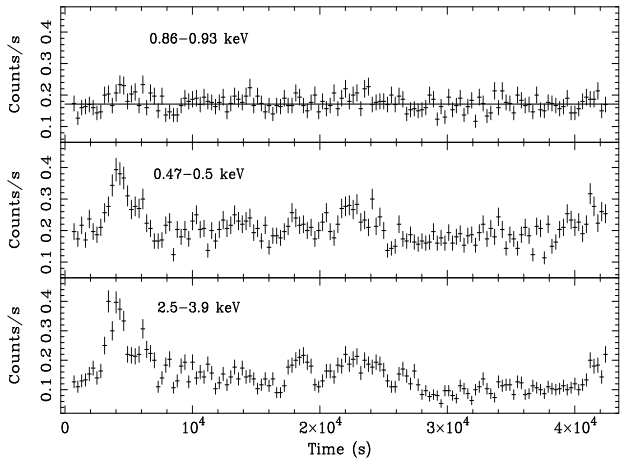} 
\end{center}

\caption{Source light curves during the rev. 541. {\it The top panel} shows the light curve in the big drop 
of variability energy band (0.86--0.93 keV, see Fig.~\ref{RMS}); {\it the middle panel} shows the same in the 
0.47--0.5 keV band; {\it the bottom panel} shows the 2.5--3.9 keV band.}
\label{lcsoft}
\end{figure}

In the bottom panel of the same Figure, we show the RMS spectrum
obtained for the low flux observation during rev~541. The variability
below 3~keV is strongly suppressed with respect to the high flux
observation. Moreover, the shape of the RMS spectrum is totally
different and rather unusual. The trend of increasing variability
toward softer energies breaks down completely and the most striking
feature is the marked drop of variability around 0.9~keV. This
feature, as well as the other drop of variability at $\sim$0.55 keV,
could have the same nature as the structure seen in the high flux RMS
spectrum, but is much more significant and prominent here. As a sanity
check, we show in Fig.~\ref{lcsoft} the source light curve around
0.9~keV and in two bands below and above for comparison. The source
appears to be constant around 0.9~keV ($\chi^2$=144.2 for 140 degrees
of freedom). The constant hypothesis is unacceptable fits in
the other two bands. As in the high flux observation, the variability
is also suppressed around 6.4~keV. The drop at 6--7~keV is here even
more dramatic due to the reduced continuum level and therefore
increased visibility of the constant narrow Fe line.

\subsection{Constant components}

The RMS spectrum analysis given above indicate the presence of two
constant or weakly variable components. The first is related to X--ray
reflection and Fe emission and is seen in both observations. The
second is associated with the drop in variability around 0.9~keV which
is clearer in the low flux data where the contribution of the
continuum is lower.

\subsubsection{Reflection from distant material} 

Fig. \ref{RMS} shows a clear minimum at 6.4~keV indicating that the
narrow Fe emission line is less variable than the continuum in both
observations. The Fe line is clearly detected in the time--averaged
spectra of both observations. In the low flux one, the line is
unresolved, has an energy of $6.42\pm 0.015$~keV, a flux of
$(1.5\pm0.2)\times$10$^{-5}$~ph~s$^{-1}$~cm$^{-2}$, and an equivalent
width of about 260~eV. The line energy and flux are consistent with
measurements obtained in the high flux observation and also with
previous data from {\it Chandra} (Collinge et al 2001) and {\it
  BeppoSAX} (Guainazzi et al 1998). As mentioned, a constant and
narrow Fe emission line is also required by the RMS spectra. Since the
Fe line must be associated with a reflection continuum, in all
subsequent fits we always include a constant and neutral reflection
model from Magdziarz \& Zdziarski (1995) and a narrow Fe line. The
reflection continuum normalization is chosen such that the Fe line
equivalent width (with respect to such continuum only) is consistent
with the {\it BeppoSAX} observation (EW$_{Fe K}\sim$700 eV) by
Guainazzi et al (1998). This is because that observation caught
NGC~4051 in a particularly low flux state in which the primary
continuum had switched off leaving only reflection from distant matter
in the hard band, thereby providing a most useful way to constrain
that component (under the reasonable assumption that it remains
constant).

\subsubsection{Photo-ionized plasma emission}

\begin{figure}
\begin{center}
\includegraphics[width=0.34\textwidth,height=0.46\textwidth,angle=-90]{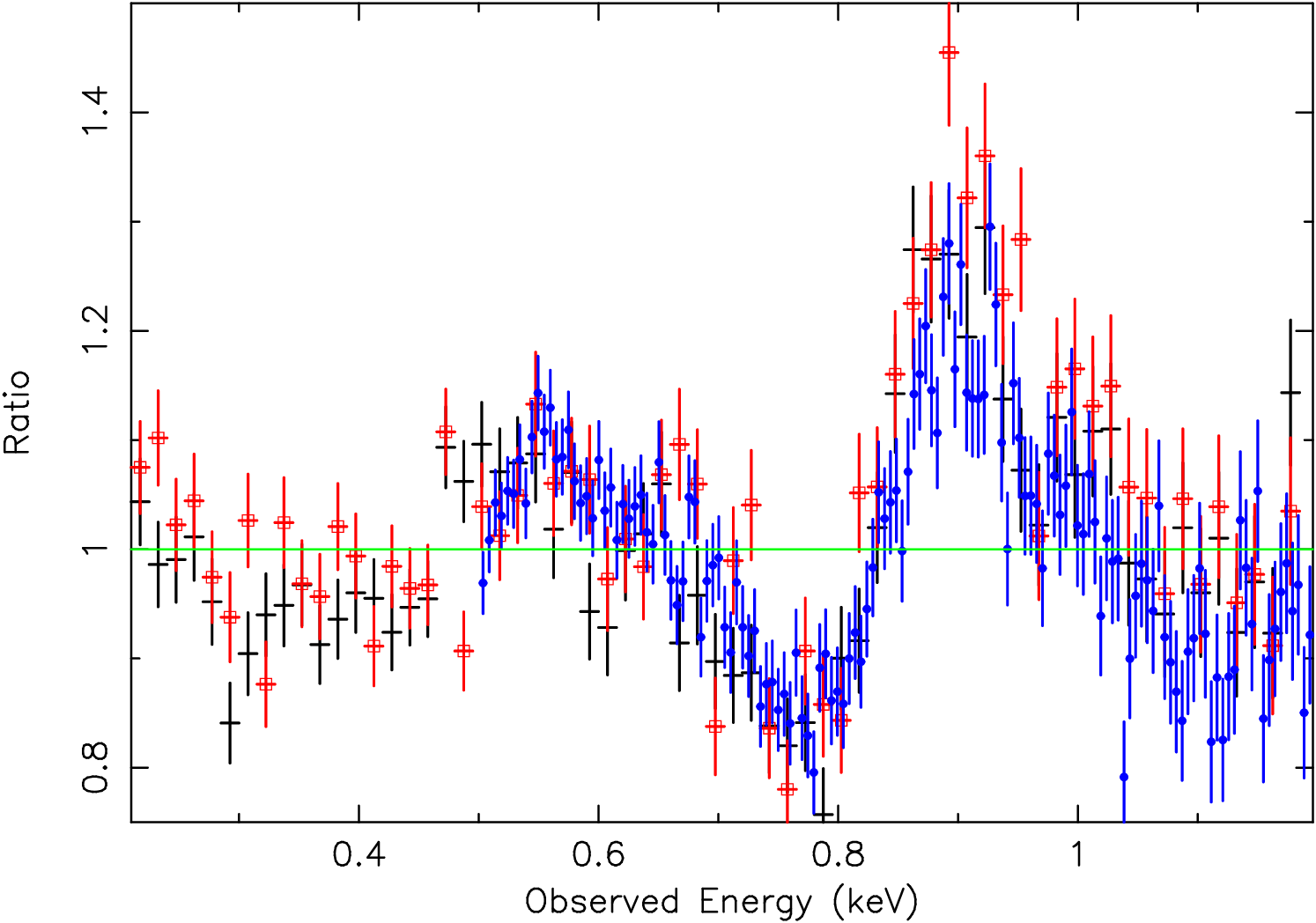} 
\includegraphics[width=0.34\textwidth,height=0.46\textwidth,angle=-90]{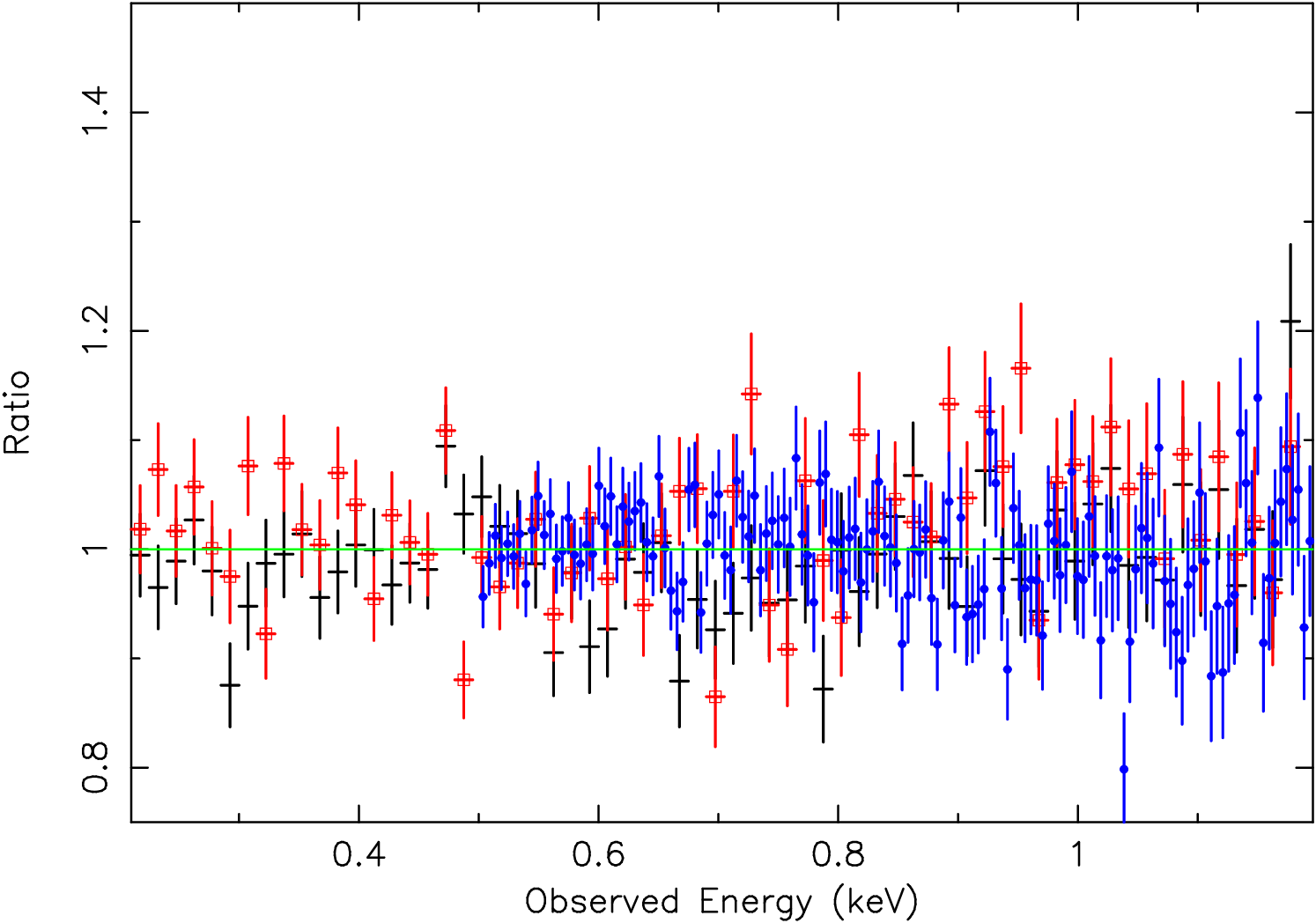} 
\end{center}
\caption{A portion of the MOS and pn data during the low flux
  observation. The top panel shows the ratio to a simple continuum
  model (power law plus thermal emission absorbed by the Galactic
  column). In the bottom panel we show the ratio to a model in which
  we added Oxygen lines (and edge) and emission around 0.9~keV
  consistently with what is seen in the high--resolution simultaneous
  RGS data.}
\label{R12}
\end{figure}

The second constant component is clearer in the low flux state RMS
spectrum (bottom panel of Fig.~\ref{RMS}) and appears as a deep
minimum, that reach almost null variability, at around 0.9~keV and
with a lower variability at $\sim$0.55 keV (see also
Fig.~\ref{lcsoft}).  These prominent drops indicate the presence of
constant emission around $\sim$0.9 and $\sim$0.5~keV, as the drop at
6.4~keV is easily interpreted as constant Fe emission.  Moreover, an
emission--like feature is seen in the pn spectrum of the low flux
observation and in previous {\it Chandra} low flux spectra as well
(Collinge et al 2001; Uttley et al 2003). The energy of the feature in
the pn spectrum is $\sim$ 0.9~keV, consistent with
Ne~{\footnotesize{IX}} (and possibly Fe emission plus
O~{\footnotesize{VIII}} recombination continuum, or RRC). A second
feature is also detected around 0.5--0.6~keV, where emission from
O~{\footnotesize{VII}} and O~{\footnotesize{VIII}} is expected. As an
illustration, in the top panel of Fig.\ref{R12}, we show the ratio of
the low flux pn and MOS data to a simple continuum model comprising
Galactic absorption power law and blackbody in the relevant energy
band. The energies of the emission features and their intensities over
the continuum correspond with the two drops seen in the RMS spectrum.
The larger strength of these features during rev. 541 is due to the
lower continuum. In fact, the same drop of variability at about 0.9
keV observed during the rev. 263 can be reproduced by the same
constant emission lines with a higher continuum.

\begin{figure}
\begin{center}
 \includegraphics[width=0.34\textwidth,height=0.46\textwidth,angle=-90]{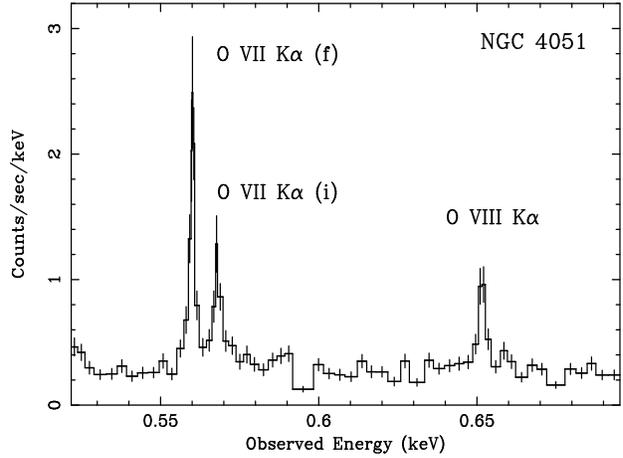} 

\end{center}
\caption{A portion of the RGS 1 spectrum of NGC~4051 in its low
  flux state. Emission by photo-ionized gas is clearly visible in the form
  of several emission lines. Here we show the O K$\alpha$ lines only.
  Such emission line spectrum is generally seen in Seyfert~2 galaxies
  and is detectable here thanks to the extremely low continuum.}
\label{rgs}
\end{figure}

To understand the nature of these constant emission components, we
then inspected the simultaneous high--resolution RGS data to better
search for possible narrow emission lines that could explain the
features seen at CCD resolution. As already shown by Pounds et al
(2004) the high--resolution RGS data during the low flux {\it
  XMM--Newton} observation indeed show an emission line
spectrum which is very similar to that of typical Seyfert 2 galaxies
(e.g.Kinkhabwala et al 2002; Bianchi et al 2005).  We take the same
continuum as in the CCD data and add a O~{\footnotesize{VII}} edge at
0.74~keV to account for the absorption seen there. The most prominent
emission lines are due to the O~{\footnotesize{VII}} triplet
(dominated by the forbidden and intercombination lines and with no
clear resonance), the O~{\footnotesize{VIII}} K$\alpha$ line and the
N~{\footnotesize{VII}} K$\alpha$ line (see Fig.~\ref{rgs}). The
Ne~{\footnotesize{IX}} (forbidden) line is also clearly detected at
0.905~keV with a flux of $(3.5 \pm 1.0) \times
10^{-5}$~ph~s$^{-1}$~cm$^{-2}$, not enough to account for the pn
feature around 0.9~keV. However, the Ne line sits on top of a broad 
feature most likely due to O~{\footnotesize{VIII}} RRC and possibly
unresolved Fe emission lines.  When fitted with a crude Gaussian model
in the RGS data, such a feature has an energy of $0.88\pm 0.01$~keV
and a flux of $(5.0 \pm 1.5)\times 10^{-5}$~ph~s$^{-1}$~cm$^{-2}$. By
combining the Ne line with this broad feature ($\sigma \simeq
20$~eV) we obtain a flux of (0.7-1.1)$\times
10^{-4}$~ph~s$^{-1}$~cm$^{-2}$ around 0.9~keV.

We then included (with all parameters fixed to the RGS results) the N
and O lines in the MOS/pn model, together with a
O~{\footnotesize{VII}} edge at 0.74~keV (fixed energy) and a Gaussian
emission line with free energy, width, and normalization to account
for the blend of Ne line plus broad feature around 0.9~keV. The
resulting fit is acceptable and is shown as a ratio plot in the bottom
panel of Fig.~\ref{R12}. As for the emission feature at 0.9~keV, we
measure an energy of $0.88 \pm 0.02$~keV and a flux of $(1.2 \pm 0.2)
\times 10^{-4}$~ph~s$^{-1}$~cm$^{-2}$, consistent with the RGS upper
limit. As mentioned in Pounds et al (2004) excess absorption might be
present around 0.76~keV (possibly related to a M--shell unresolved
transition array from Fe). If included in both CCD and RGS data, this
has the effect of reducing the broad 0.9~keV feature intensity,
making so much more difficult to reproduce the variability drop in the
RMS spectra.

The above analysis demonstrates that emission from photoionized gas is
present in the low flux data and detected both in the RGS and in the
lower resolution CCD detectors. Such components, as shown by the RMS
spectra, are constant over time, thus then are likely to come
from relatively extended gas (see Pounds et al 2004). A constant
emission component, indeed, does explain the unusual drop of
variability around 0.9~keV in the low flux RMS spectrum. The line
intensities are also consistent (though not detected) with the higher
flux observation data.  They are not detected because of the much
higher continuum level. We then included, in addition to the constant
neutral reflection from distant material, such lines with fixed
energies and intensities (as obtained from the low flux RGS data) in
all subsequent fits.

\section{Flux--flux plots: is pivoting the only solution?}

Another model--independent approach that provides information on the
dominant driver of the spectral variations is the flux--flux technique
(see Taylor, Uttley \& M$^{\rm c}$Hardy 2003; Uttley et al
2004).  We recall here that a linear flux--flux relationship reveals
that the data can be described by a simple two--component model where
one component varies in flux but not in spectral shape, while the
other remains constant both in flux and spectral shape. An offset on
the hard or soft axis allows one to infer in which of the two bands
the constant component contributes the most.  On the other hand, if
the flux--flux relationship has a power law functional form, then the data
are consistent with spectral pivoting of the variable
component.

  Uttley et al (2004) plotted the 2--10~keV band count rate versus the
  0.1--0.5~keV one and found a very accurate power law flux--flux
  relationship which was interpreted as evidence for spectral pivoting
  of the variable component. 
  The choice of the soft and hard energy ranges by Uttley et al (2004)
  is driven by the requirement of having well separated bands to best
  reveal the spectral variability. However the 0.1--0.5~keV band does
  not seem to have the same variability properties of the broadband
  continuum. This is clearly seen in the top panel of Fig.~\ref{RMS}
  in which the soft plateau indicates the presence of a component with
  different variability properties than the continuum above 1~keV. Such
  a component is likely to be that producing the soft excess.
  Therefore we consider as the reference soft band the 1--1.4~keV
  range to avoid possible contamination from undesired components. In
  all previously reported analysis of the X--ray spectrum of NGC~4051,
  the intermediate 1--1.4~keV band is dominated by a power law
  continuum which is not strongly affected by absorption, soft excess,
  or reflection.  As for the hard band, we consider the 4-10~keV
  range, simply to allow enough leverage for spectral variability to
  show up clearly.  

\begin{figure}
 \includegraphics[width=0.34\textwidth,height=0.46\textwidth,angle=-90]{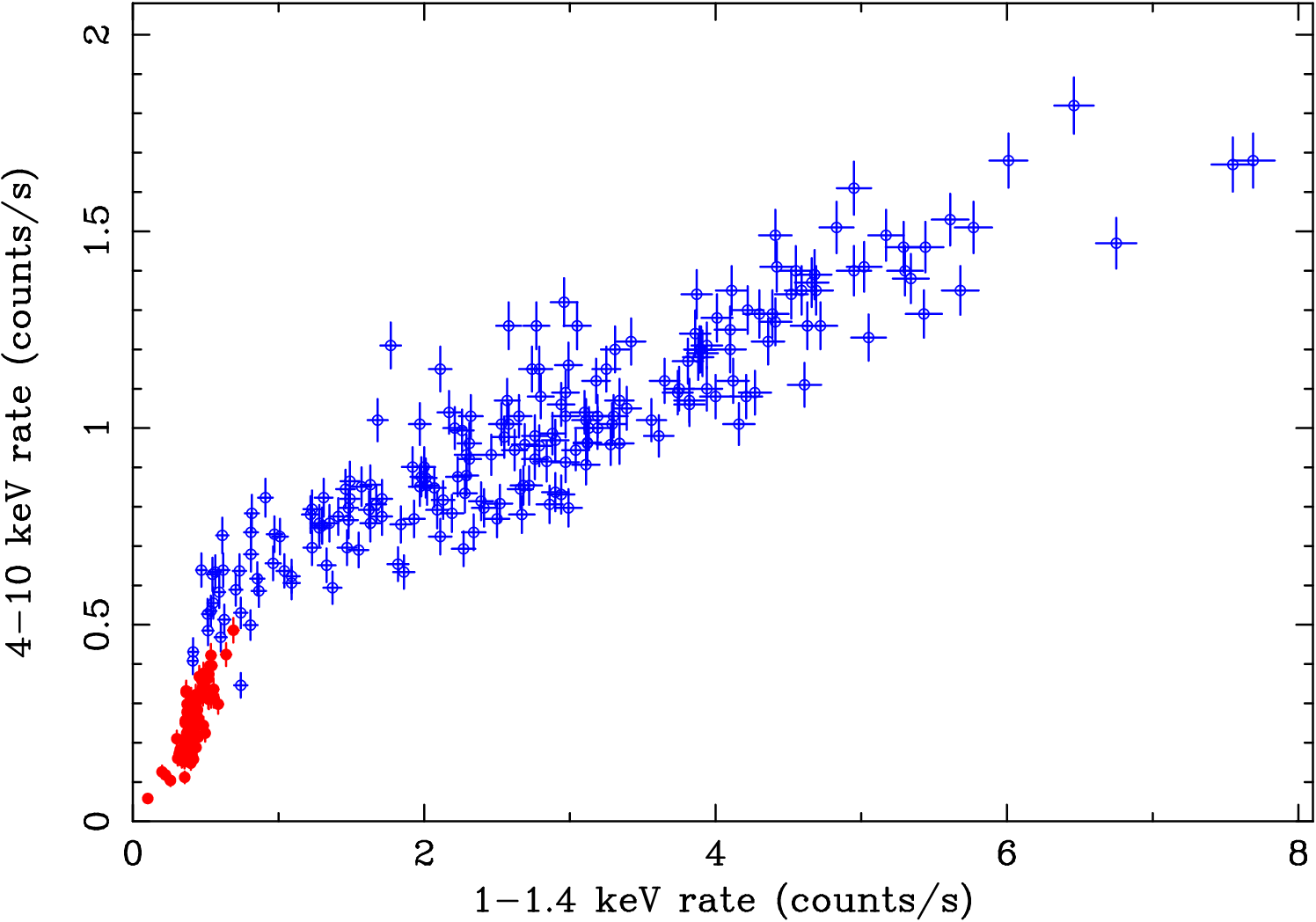}
 \includegraphics[width=0.34\textwidth,height=0.46\textwidth,angle=-90]{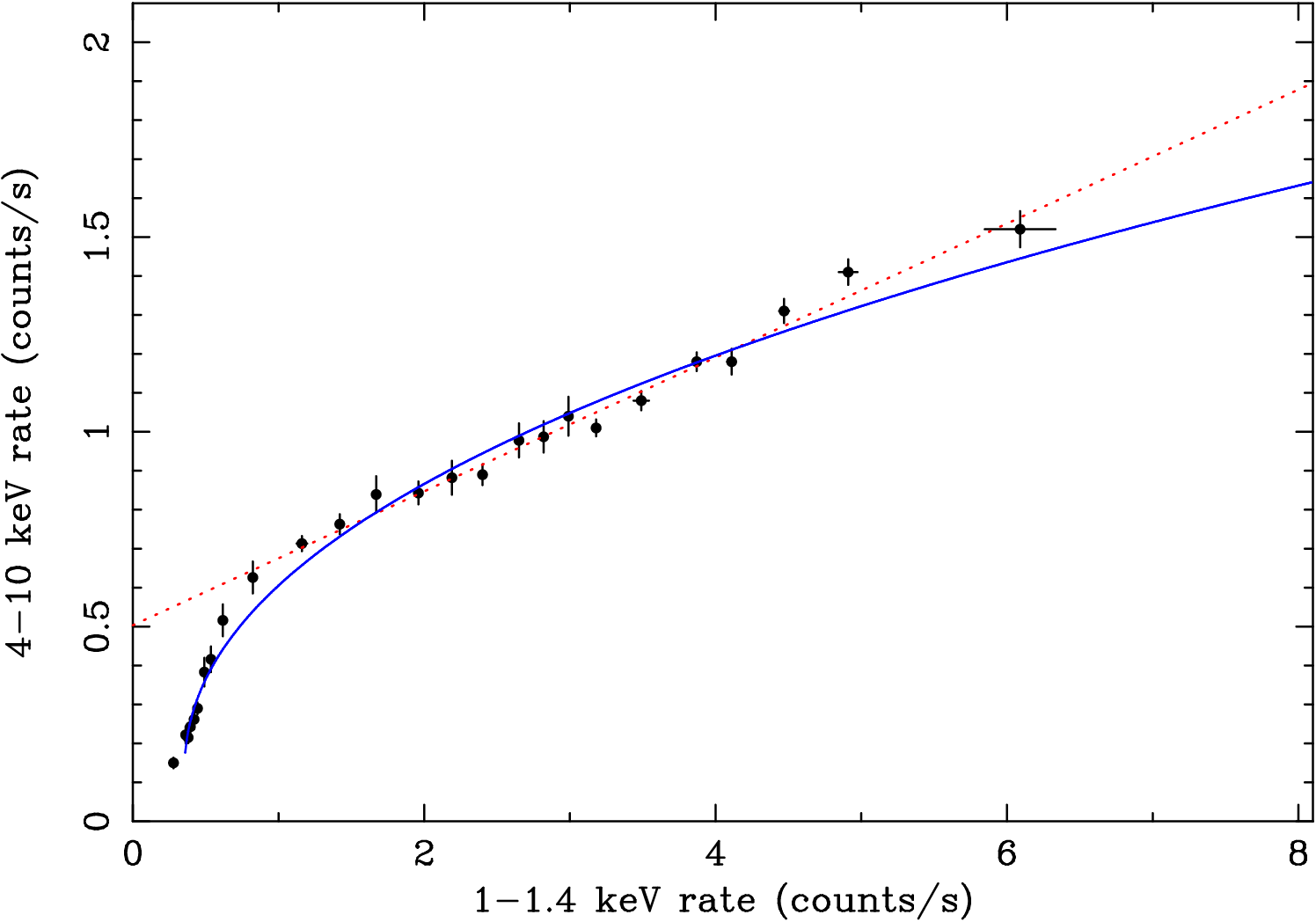} 
\caption{The flux--flux plot obtained by plotting the (pn) 4--10~keV
  count rate versus that in the 1--1.4~keV band. The bottom panel is
  just the binned version of the original flux--flux plot (top panel).
  The solid line is the best fit power--law relationship, while the
  dotted one is the best fit linear relationship if the lowest flux
  data points are excluded (see text for details).}
\label{ff1}
\end{figure}

  In the top panel of Fig.~\ref{ff1}, we show the flux--flux
  relationship in the chosen hard and soft bands. The two {\it
    XMM--Newton} observations are shown with different symbols.  As
  already pointed out by Uttley et al (2004) the low flux state data
  join gently on to the normal/high flux state ones, showing that the
  spectral variability is likely to be due to the same mechanism at
  all flux levels.  There is no dramatic discontinuity between the two
  observations suggesting, once again, that the spectral changes are
  not due to a dramatic change in one of the physical parameters (such
  a substantial change in the absorbing column, as suggested by Pounds
  et al 2004).  The flux--flux relation is clearly non--linear with
  some curvature showing up especially at low fluxes.  In the bottom
  panel, we show a binned version of the same plot (minimum of 10
  points per bin).  In the same figure, the solid line is the best fit
  for a power law model representing a situation in which the main
  driver of the spectral variability is spectral pivoting (see e.g.
  Taylor, Uttley \& M$^{\rm c}$Hardy 2003 for details).  The fit
    is performed as in Uttley et al (2004) with a power law
    relationship including constant components which are required in
    both the hard and the soft band.
  The power law relationship does not provide a good fit to the data
  ($\chi^2=62.4$ for 22 degrees of freedom), mainly because the model
  has too much curvature at high fluxes and not enough at low fluxes.
  This does not exclude spectral pivoting as the main driver for the
  spectral variability, but suggests that we need to look for an
  alternative explanation.

In fact, if the lowest flux data points are excluded, the flux--flux
relationship is linear with high accuracy. This is shown as a dotted
line in the bottom panel of Fig.~\ref{ff1} which is the best fit
linear relationship
obtained when the lowest flux data points are ignored. We obtain a
very good fit ($\chi^2=14.3$ for 15 degrees of freedom) with
large hard offset of $0.50\pm
0.02$. As mentioned, a linear relationship suggests that the spectral
variability is due to a two--component model, i.e. to the relative
contribution of a variable component whose spectral shape is
flux--independent (e.g. a constant $\Gamma$ power law varying in
normalization only) and a constant or weakly variable harder component. If
the flux--flux relationship at normal/high fluxes is indeed linear,
one can compute the contribution of the constant component in the
4--10~keV from the y--axis offset $C_{\rm h}$. Since the mean
4-10~keV rate for the normal/high flux data points is $F_{\rm h}
\simeq 1.0$, this means that the putative constant component
contribution in the 4--10~keV band is as large as 50 per cent in the
normal/high flux states.  

However, if indeed the two--component model
is a fair representation of the spectral variability at normal/high
fluxes, the low flux data imply that the spectral variability must
smoothly change regime at low fluxes producing the gentle low--flux
bending in the relationship. Such a transition could be reproduced
if, below a given flux level, the constant component (or part of it)
is not constant anymore but rather follows the changes of the variable
one. In fact, any two--component model in which the less
variable component varies together with the variable one at low
fluxes and saturates to an approximately constant flux at higher
fluxes would produce the same kind of relationship as observed here.

\subsection{Is the two component model consistent with the shape of
  the RMS spectrum?}

  In order to check if the source spectral variability could be due
  to the two--component model we try to reproduce the RMS spectrum
  during rev. 263 with a simple phenomenological parametrization (the
  RMS spectrum during rev.  541 is not considered because its shape is
  mainly dictated by the constant components coming from distant
  material).  The model is composed of a direct Power--Law Component
  (PLC) and an ionized Reflection--Dominated Component (RDC, from Ross
  \& Fabian 2005) the spectrum of which is convolved with a
  {\small{\tt LAOR}} kernel with fixed inner and outer disc radius (to
  $1.24~r_g$ and $100~r_g$ respectively), and fixed disc inclination
  ($30^\circ$).  In the RMS simulation the contribution from neutral
  distant reflection is also considered.

  We assume that all the source variability is due to the variation of
  the normalisation of the direct PLC that has a spectral index fixed
  at $\Gamma$=2.4 (Lamer et al.  2003) and that the reflection
  component is completely constant not only in normalisation, but also
  in spectral shape (i.e. ionisation and relativistic blurring
  parameters). To simulate the RMS we vary the PLC normalization from
  the lowest to the highest value observed. For the RDC we have chosen
  the best fit values of the spectrum during the low flux state of
  rev.  263 ($\xi \simeq$50 and disc emissivity index of
  $\alpha\simeq$6). Figure 8 shows this simulated RMS (dotted line,
  green in the colour version) and, for comparison, the simulated RMS with the
  reflection parameters of $\xi$=5 and $\alpha$=4 (solid line, blue in
  the colour version), that also roughly reproduce the low flux spectrum.
\begin{figure}
 \includegraphics[width=0.34\textwidth,height=0.46\textwidth,angle=-90]{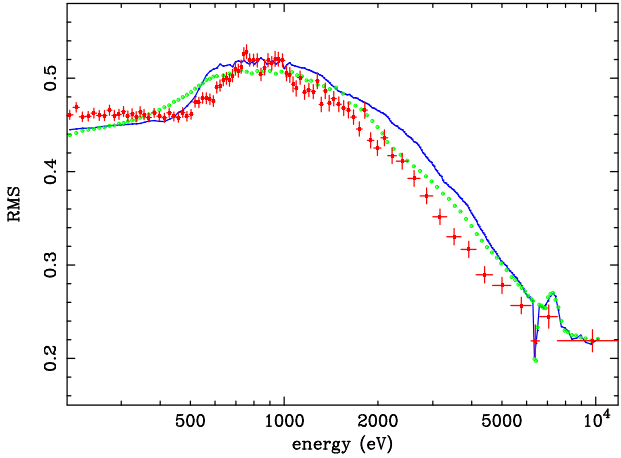}
 \caption{The observed RMS spectrum obtained during rev. 263 is
   compared with synthetic ones. The solid line (blue in the colour
   version) is obtained varying in normalization (from the lowest to
   the highest value observed) a PLC with $\Gamma$=2.4 and keeping a
   RDC with ionization parameter $\xi$=5 and disc emissivity index
   $\alpha$=4 completely constant.  While the simulated RMS shown by
   the dotted line (green in colour) is calculated in the same way as
   above, but with $\xi\simeq$50 and $\alpha\simeq$6. In both models the
   contribution from neutral distant reflection is considered. }
\label{simRMS}
\end{figure}
The interplay between the PLC and the RDC can roughly reproduce the
shape of the RMS (within $\Delta RMS \lsimeq$0.02).  This means that
the two component model can explain the bulk of the spectral
variability requiring no further variable component at low energy.
The variability in the soft band can be explained by a variable power
law and a constant soft excess (which is in this case due to constant
RDC)

Clearly, a completely constant reflection component is an
over--approximation. In fact, although, it is possible to reproduce
the bulk of the variability, some variations of the normalization of
the RDC are required in order to match with more details the RMS
spectra and to reproduce the constant spectral shape below 0.5~keV
(see Section 6.3.2).  

\section{Time--resolved spectral analysis}

In sources such as NGC~4051 where large amplitude fast variability is
associated with spectral variability, time--resolved or flux--resolved
spectroscopy provides much more detailed information on the nature of
the source than time--averaged spectroscopy. In the particular case of
NGC~4051, flux--resolved spectroscopy could be a risky procedure to
adopt because of the small central black hole mass (and therefore the
short dynamical timescale associated with it).  As an example, if we
were to consider the very high flux state of the first {\it
  XMM--Newton} observation (see left panel of Fig.~\ref{lcs}) and make
a selection say above 40 counts/s, we would accumulate spectra from
intervals sometimes spaced by 70~ks. This timescale is about 140 times
the dynamical timescale at 10~$r_g$ (where $r_g=GM/c^2$) for a black
hole mass of $\sim 5\times 10^5~M_\odot$ (e.g. Shemmer et al 2003)
which is, in our opinion, too large a factor to extract any truly
significant information on the nature of the spectral variability of
the source.

We therefore explore time--resolved spectroscopy on the shortest
possible timescale (set by requiring good quality time--resolved
individual spectra). In the following, we present results obtained by
performing such an analysis on a 2~ks timescale (about four dynamical
timescales at 10~$r_g$). The 2~ks slices that have been used in the
analysis are shown in Fig.~\ref{lcs} and have been numbered for
reference.

\subsection{The 2--10~keV $\Gamma$--flux relationship}

\begin{figure}
 \includegraphics[width=0.34\textwidth,height=0.46\textwidth,angle=-90]{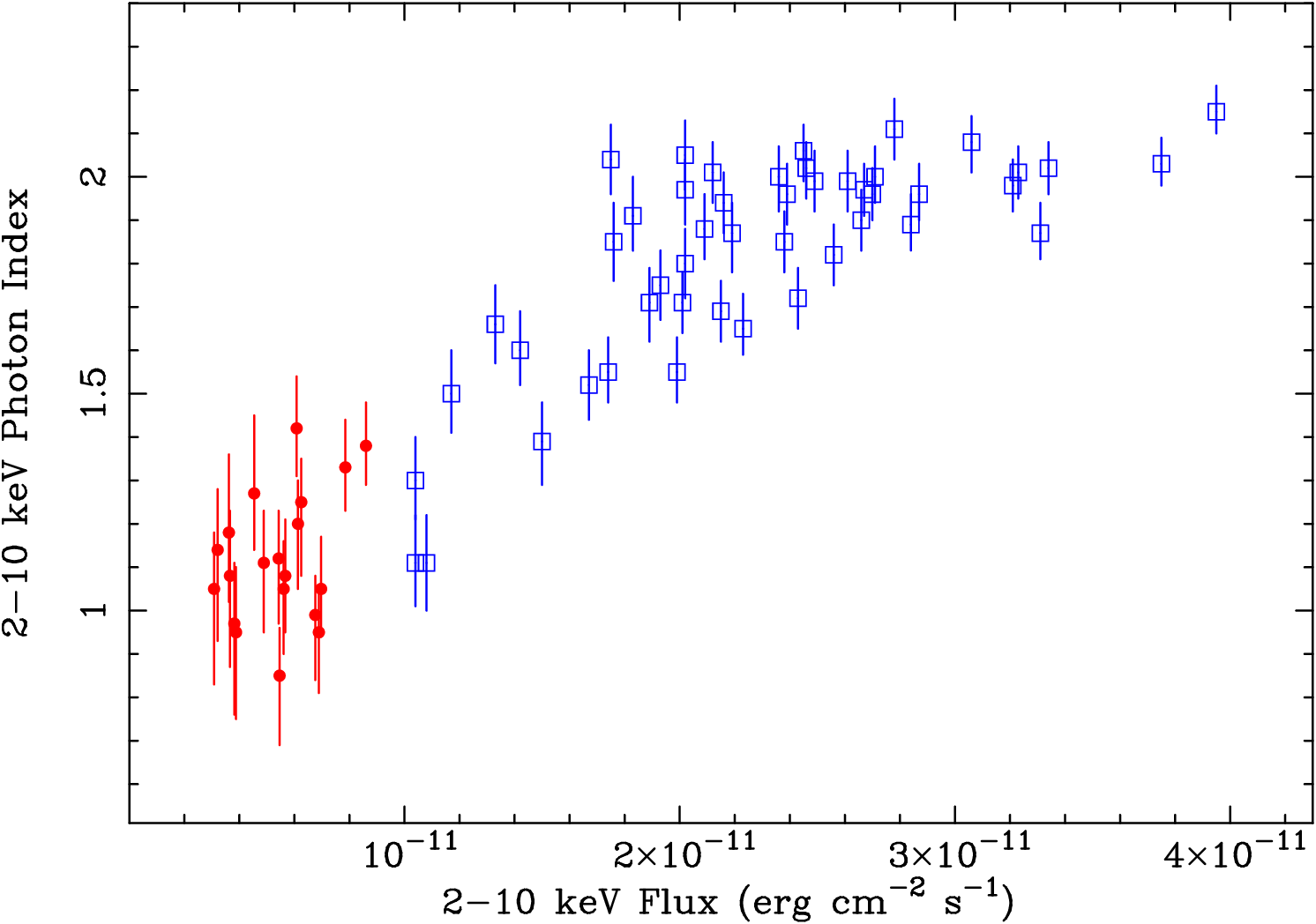}
 \includegraphics[width=0.34\textwidth,height=0.46\textwidth,angle=-90]{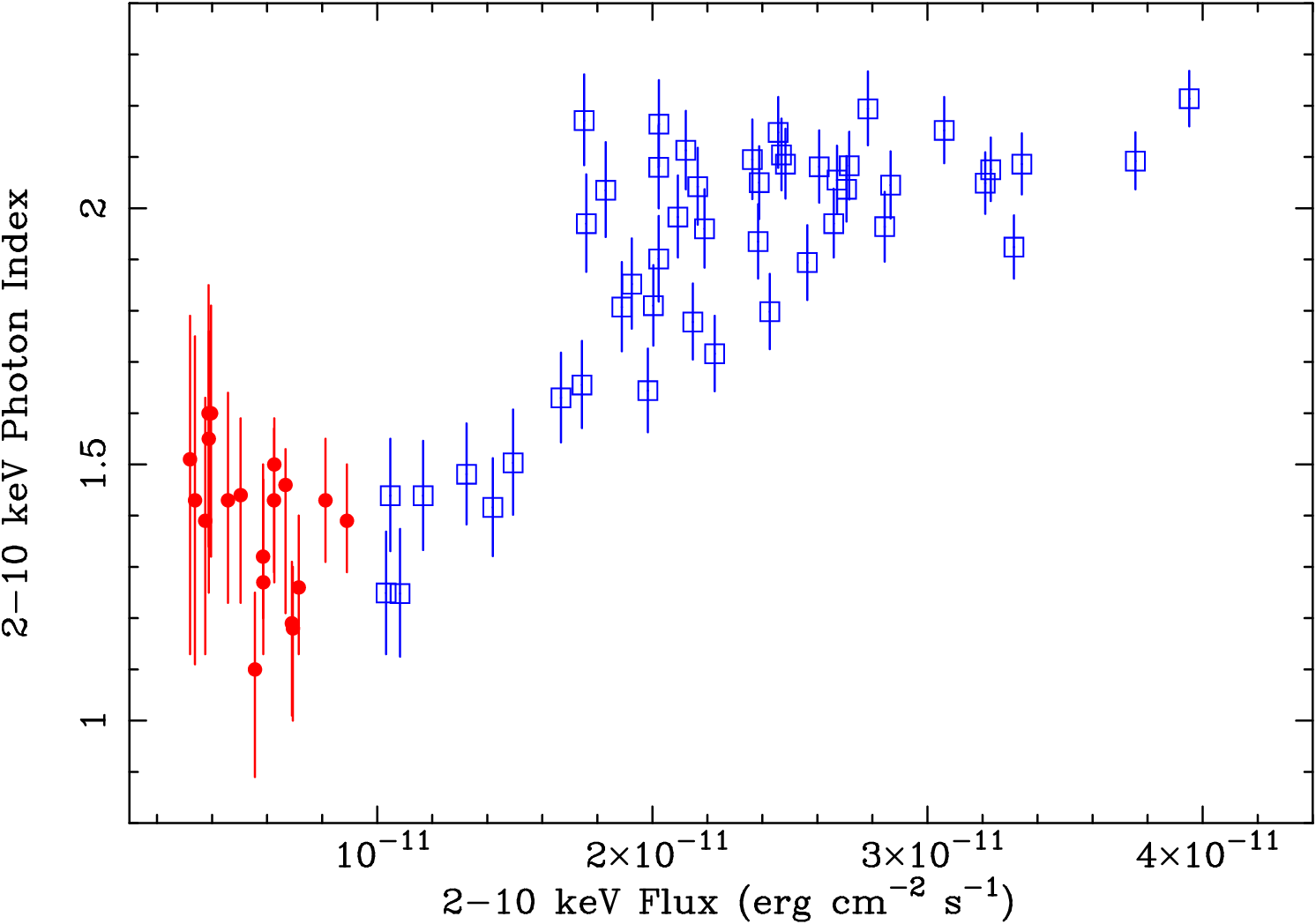} 
\caption{The best fit spectral index vs. flux in the 2--10 keV
  band. In the top panel the spectral model is a simple power law,
  while constant reflection (including a narrow Fe line) is added to
  the model for the bottom panel results.}
\label{gf1}
\end{figure}

It is well known that the 2--10~keV spectral slope in NGC~4051 (and
other sources) is correlated with the source flux. In the top panel
of Fig.~\ref{gf1}, we show such a correlation when all the 2~ks spectra are
fitted with a simple power--law model in the 2--10~keV band. Data for
the two observations are shown with different symbols. The photon index
increases with flux and seems to saturate around $\Gamma\simeq 2.2$
with a $\Delta\Gamma\simeq 1.2$ between the low and high flux states.
However, the power law model is clearly inadequate to fit the
2--10~keV spectra in NGC~4051. This is because of the presence of a
narrow 6.4~keV Fe K$\alpha$ line and the associated reflection
continuum in the hard band. As shown above, such a component is constant
and likely originates in some distant material (such as the torus).

We then fit again the 2~ks data by including the constant reflection
component and show in the bottom panel of Fig.~\ref{gf1} the resulting
$\Gamma$--flux relationship. The $\Delta\Gamma$ between the low and
high flux states is now reduced to about 0.7--0.8.  Again the
  photon index seems to saturate at high fluxes to $\Gamma\simeq 2.2$.
  This value is slightly different from the one ($\Gamma\simeq
  2.3-2.4$) observed during a long RXTE observation campaign (Lamer et
  al. 2003). The difference could be due to long term spectral
  variability.  Moreover, the photon index seems to saturate not only
at high fluxes but also at low fluxes (to $\Gamma=1.3$--$1.4$).  Since
the two asymptotes are not extremely well defined, such behaviour
could still be consistent with spectral pivoting, even though it
should be stressed that the low flux photon index is still somewhat
hard to be explained in terms of standard Comptonization models and
would suggest some peculiar physical scenario (such as a
photon--starved corona).

On the other hand, the $\Gamma$--flux relationship seen in the bottom
panel of Fig.~\ref{gf1} can be
explained if an additional and weakly variable component is present in
the 2--10~keV band. If so, such a component dominates the low flux
states and the hard photon index measured there is just a measure of
its intrinsic spectral shape in the 2--10~keV band, while it is
overwhelmed by the power law at high fluxes where $\Gamma\simeq 2.2$
is then the intrinsic power law photon index.

In other words, as an alternative to spectral pivoting, the
$\Gamma$--flux relationship can be explained by assuming the presence
of i) a constant reflection component from distant matter; ii) a
variable power law with constant or weakly variable slope
$\Gamma\simeq 2.2$; iii) an additional component with approximate
spectral shape of $\Gamma= 1.3$--$1.4$ in the 2--10~keV band which
must be almost constant at normal/high fluxes and is allowed to vary
significantly only at low fluxes. 

We point out here that our flux--flux plot analysis showed that the
two--component model might indeed be appropriate in the normal/high
flux states. If so, the constant component contributes by about 50 per
cent in the 4--10~keV band (from the y--axis offset in the linear fit)
and this is much too large a fraction to be accounted for by constant
reflection from distant material (and associated narrow Fe line) which
contributes in that band by only 10 per cent.  The two--component
model interpretation of the flux--flux relationship at normal/high
fluxes implies that an additional constant component is present in the
normal/high flux states and contributes by about 40 per cent to the
4--10~keV band. Such a component naturally produces the observed
$\Gamma$--flux relationship because as the flux drops the spectrum
becomes more and more dominated by it. 

\subsection{Broadband analysis I: the standard model}

As a first step, we consider a simple continuum model comprising a
power law plus black body (BB) emission to model the prominent soft
excess. This is, in many respects, the ``standard model'' to fit AGN
X--ray spectra and, though often crude and phenomenological, provides
useful indications that can guide further analysis. The power law
slope is free to vary to account for possible spectral pivoting. We
add to the model neutral photoelectric absorption with column density
fixed to the Galactic value. To account for the presence of absorbing
ionized gas in the line of sight we include two edges
(O~{\footnotesize{VII}} and O~{\footnotesize{VIII}}).  We searched for
possible variations of their energies without finding any evidence, so
we fixed the absorption energies to the rest--frame ones.  This is
clearly a crude approximation for the effects of the warm absorber in
NGC~4051, but the quality of the 2~ks spectra is not such to suggest
the use of more sophisticated models. In addition, the constant
components discussed above (reflection from distant matter and
emission from photoionized gas) are also included in the spectral
model. The overall model has six free parameters only.

The resulting fits are very good with a reduced $\chi^2$ ranging
between 0.9 and 1.2 with a mean close to unity. Therefore, from a
statistical point of view, a single model is
able to describe the spectral variability of the source in a very
satisfactory way. Below, we discuss the main results of the fitting
procedure and their main physical consequences.

\subsubsection{Variations of the central source}

\begin{figure}
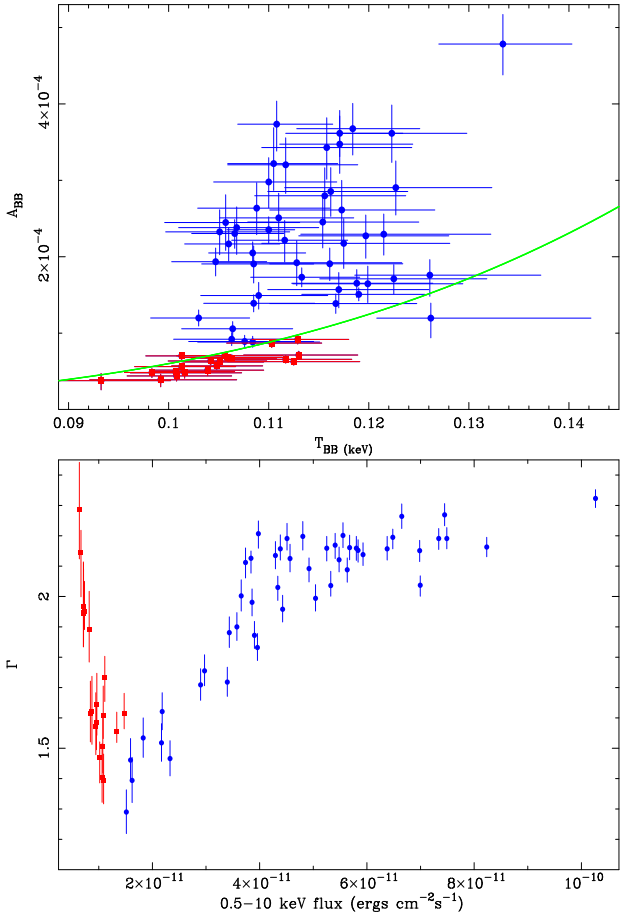

 \includegraphics[width=0.34\textwidth,height=0.46\textwidth,angle=-90]{TBB_ABB_2edgesMOS.cps}
 \includegraphics[width=0.34\textwidth,height=0.46\textwidth,angle=-90]{Gamma_Flux_2edges.cps} 
 \caption{In the top panel we show the blackbody intensity as a
   function of its observed temperature. The green line shows the best
   fit relation between the A$_{BB} \propto$T$_{BB}^4$. In the bottom
   panel, we show the broadband $\Gamma$--flux relationship as
   obtained from the standard model fits (see text for details).}
\label{BBpo}
\end{figure}

In Fig.~\ref{BBpo} (top panel) we show the relation between the
measured BB temperature and its intensity, which is proportional to
the luminosity (in the present case A$_{BB}$=10$^{-4}$ correspond to
L$_{BB}$=9.7$\times$10$^{40}$ ergs s$^{-1}$, assuming H$_{0}$=71 Mpc
km$^{-1}$ s$^{-1}$). The BB luminosity spans about one order of
magnitude, while the BB temperature ranges from about 90~eV to 140~eV
with an average of 110--120~eV. Such a temperature is somewhat high,
but can still be consistent with that expected from a black hole with
mass $\simeq 5\times 10^5~M_\odot$ (Shemmer et al 2003;
M$^{\rm{c}}$Hardy et al 2004) accreting at high rate (for the given
black hole mass, the expected disc temperature is of the order of
70~eV for an object accreting at the Eddington rate). We point out
that even in the lowest flux states the temperature (about 100~eV) is
larger than predicted and, in the framework of standard disc models,
would already require the system to be accreting at the Eddington
rate.

If the soft excess in NGC~4051 is really due to BB emission from a
constant area, we should be able to detect the BB law (L$_{BB} \propto
T^4$) in data spanning one order of magnitude in BB luminosity.
However, such a relation is ruled out. A fit of the L$_{BB}$ versus
temperature data (top panel of Fig.~\ref{BBpo}) with the BB law
(leaving only the normalization as a free parameter) produces a very
bad fit with $\chi^2 = 1552$ for 67 degrees of freedom. The relation
appears to be much steeper (with an index of 6.3--6.4 rather than 4),
but even so, a power law fit is unacceptable.  The measured
temperature appears to be much more constant (the fit with a constant
gives a $\chi^2 = 75$ for 67 degrees of freedom) than predicted by
standard disc models.

The properties of the soft excess in NGC~4051 are remarkably similar
to those of 26 bright radio--quiet quasars (considered as a
population) studied e.g. by Gierlinski \& Done (2004) and selected
from the {\it{XMM--Newton}} archive by their strong UV continuum (such
that they have strong accretion disc emission). The quasar sample
spans a wide range of black hole masses and luminosities and should
therefore exhibit a wide range of disc temperatures (between a few eV
to about 70~eV). However, when fitted with a cool Comptonized
component, the measured temperature of the soft excess is remarkably
constant throughout the sample with a mean of 120~eV (and very small
variance). It is a rather surprising coincidence that this is exactly
the average temperature we measure for the soft excess in NGC~4051,
considering also that the black hole mass in NGC~4051 is about 3
orders of magnitude smaller than the typical black hole mass in the
quasar sample. Moreover, no correlation between the measured soft
excess temperature of quasars and the expected disc temperature was
found, in line with the results presented above.

Such discrepancies naturally raises questions on the real
nature of the X--ray soft excess. As pointed out by Gierlinski \& Done
(2004), the remarkable constancy of the ``soft excess temperature''
might indicate an origin in atomic rather than thermal physical conditions.
Possible candidates seems to be absorption (e.g. from a disc wind, as
suggested by Gierlinski \& Done 2004) and/or ionized reflection (e.g.
from the accretion disc itself, as suggested e.g. in Ross \& Fabian
2005).

The ``standard model'' of the source emission seems to be inadequate
to describe the corona parameters as well. In the bottom panel of
Fig.~\ref{BBpo} we show the power law slope as a function of the
0.5--10~keV flux. In the normal/high flux observation the photon index
steepens with flux, as already pointed out in the 2--10~keV
analysis. This behaviour is consistent with spectral pivoting.
Once again, the lowest $\Gamma$ are of the order of 1.4 which is a
hard spectral shape to be accounted for by standard Comptonization
models and might require a photon--starved corona scenario. The main
problems arise at very low fluxes, where
the $\Gamma$--flux relation is inverted in a manner
that is not consistent with simple spectral pivoting. Such behaviour
could indicate the presence of an additional soft (steep) component
which becomes prominent at low fluxes and is not properly accounted
for by the black body component. In fact, the power law is trying to
fit the soft data by steepening the index at low fluxes and leaves
significant residuals in the hard band where a slope of about 1.3--1.4
would be more appropriate even at very low fluxes (see
Fig.~\ref{gf1}).

As a final comment, the mean optical depth of the ionized absorber
(here modelled crudely by the O~{\footnotesize{VII}} and
O~{\footnotesize{VIII}} edges) is of the order of 0.1--0.3, 
  consistent with the value obtained from the RGS analysis (Pounds et
  al. 2004).  These values are left free to vary and they seem to
suggest some variations during the two observations. Nevertheless, the
uncertainty associated with these is so big that a fit to the optical
depth with a constant during the first observation results in a
$\chi^2$ of 26 (O~{\footnotesize{VII}}) and 47
(O~{\footnotesize{VIII}}) for 47 degrees of freedom, while during the
second low flux observation the $\chi^2$ are 29 and 7 for 19 degrees
of freedom.

\subsection{Broadband analysis II: the two--component model}

The results of the fits with the ``standard model'' presented above
suggest that the soft excess is not (or not completely) truly disc
black body emission. Moreover, if the spectral variability is due to
spectral pivoting, the power law slope at low fluxes is somewhat too
hard to be accounted for by standard Comptonization (Haardt et al.
1997).  In addition, a steeper component appears in the soft band at
the lowest flux levels.  We then explore here an alternative model to
describe the spectral variability of NGC~4051.

\subsubsection{Motivation: the light bending model}

In order to reproduce the flux--flux plot and the $\Gamma$--flux
relationship without imposing spectral pivoting, we need two main
continuum components. The first one is variable in normalization only
retaining an approximately constant spectral shape at all flux levels.
We make here the assumption that this can be represented by a power
law component (PLC).  Its slope should be similar to the high--flux
asymptote of the $\Gamma$--flux relationship, i.e.  $\simeq$~2.2.  The
other component must be almost constant in normal/high flux states
where its contribution to the 4--10~keV band is of the order of 40 per
cent. In the lower flux states it has instead to correlate with the PLC in
order to reproduce the flux--flux plot. From the 2--10~keV
$\Gamma$--flux relationship we have also some indication that its
spectral shape in the hard band can be approximated by a slope of
$\Gamma=$1.3--1.4.

The obvious candidate is a relativistically blurred
reflection--dominated component (RDC) from a ionized accretion disc.
Such a component is characterised by an X--ray spectrum which is hard
in the hard band ($\Gamma=$1.3--1.6 depending on the parameters) and
steep in the soft band therefore naturally producing a soft excess. As
pointed out e.g. by Ross \& Fabian (2005) when fitted with a thermal
model in the {\it{XMM--Newton}} band, such a soft excess has a
temperature of the order of 150~eV, very similar to the constant
temperature of radio--quiet bright quasars and to the one we have
measured in NGC~4051. 

Since the RDC is produced in the inner disc by irradiation by the
PLC, the two components should always be well correlated while, in the
present case, the flux--flux plot would require that the RDC is well
correlated with the PLC at low fluxes but is then only weakly variable
at higher fluxes despite large changes in the PLC flux. This is
however precisely the main feature of the gravitational light bending
model proposed by Miniutti et al (2003) and Miniutti \& Fabian (2004).
In this model the variability is dominated by a power law component
(PLC) which changes in flux but not in spectral shape. The PLC is
(naively) assumed to have constant intrinsic luminosity and all the
PLC variability is due to light bending effects as the primary PLC
sources have different positions at different times with respect to
the central black hole. The PLC illuminates the accretion disc giving
rise to a reflection component which is affected by the relativistic
effects arising in the inner disc. Such a reflection--dominated
component (RDC) does not respond trivially to the PLC variability
because of strong gravity effects (see Miniutti \& Fabian 2004 for more
details). At low fluxes the RDC is indeed well correlated with the
PLC, but it gradually varies less and less as the flux increases
reaching a regime in which it becomes almost constant in a range in
which the PLC varies up to a factor 5. Therefore, the model
successfully reproduces the two--component model (variable PLC plus
constant RDC) at mean/high fluxes and since it predicts a correlation
between the two components at low fluxes, it has the potential of
explaining the curved flux--flux plot relationship without invoking
spectral pivoting.

In this framework, low flux states correspond to situations in which
the PLC originates few $r_g$ from the black hole strongly illuminating
the innermost disc. Light bending reduces at the same time the
observed PLC flux at infinity. As the PLC source is further away from
the black hole, more photons can escape to infinity (so that the
observed PLC flux increases) and the illuminating flux on the disc is
reduced. The model therefore predicts also a relation between the
emissivity profile of the RDC on the disc and flux. The lower the
flux, the more centrally concentrated the disc irradiation and the
steeper the emissivity.

\subsubsection{Spectral fitting results and discussion}

The two--component model we have tested comprises then i) a PLC with
fixed slope at $\Gamma=2.2$ (consistent with the high--flux asymptote
of the $\Gamma$--flux relationship) and variable normalization and ii)
an ionized RDC (from Ross \& Fabian 2005) the spectrum of which is
convolved with a {\small{\tt LAOR}} kernel with fixed inner and outer
disc radius (to $1.24~r_g$ and $100~r_g$ respectively), and fixed disc
inclination ($30^\circ$). The inner disc radius corresponds to the
innermost stable circular orbit around a Kerr black hole. The only
free parameters of the relativistic kernel is then the index $q$ of
the disc emissivity profile ($\epsilon=r^{-q}$). The ionized
reflection model is appropriate for solar abundances and has
ionization parameter and normalization as free parameters, while the
photon index of the illuminating power law in the RDC model is tied to
that of the PLC and therefore fixed to $\Gamma=2.2$. As for the
``standard model'' discussed in the previous Section, the overall
spectral model also includes Galactic absorption, constant emission
from photoionized gas and from a distant reflector and the
O~{\footnotesize{VII}} and O~{\footnotesize{VIII}} edges with fixed
energies. The number of free parameters in the model is just the same
as in the ``standard model'' case (6 free parameters).

The model reproduces very well the data at all flux levels with a
reduced $\chi^2$ between 0.8 and 1.2. Given that the standard model
and the two--component one have the same number of free parameters, we
can directly compare the resulting averaged reduced $\chi^2$ which are
$1.08\pm 0.07$ for the standard model and $1.06\pm 0.08$ for the
two--component one. Therefore, the two--component model is
statistically indistinguishable from the standard one and has to be
considered as a possible alternative to be accepted or rejected on a
physical rather than statistical basis.

As mentioned, the PLC used to model the data has a fixed slope of
$\Gamma=2.2$ at all flux levels. In fact, we also tested a variable
$\Gamma$ fit to the data and found that all 68 spectra are consistent
with $\Gamma=2.2$ with only three exceptions in which the statistics
is significantly better with a different power law slope (with small
$\Delta\Gamma < 0.15$). Therefore, we demonstrate here by direct
spectral fitting that the curvature in both the flux--flux plot and
the $\Gamma$--flux relationship does not imply spectral pivoting but
is only consistent with it.  

\begin{figure}
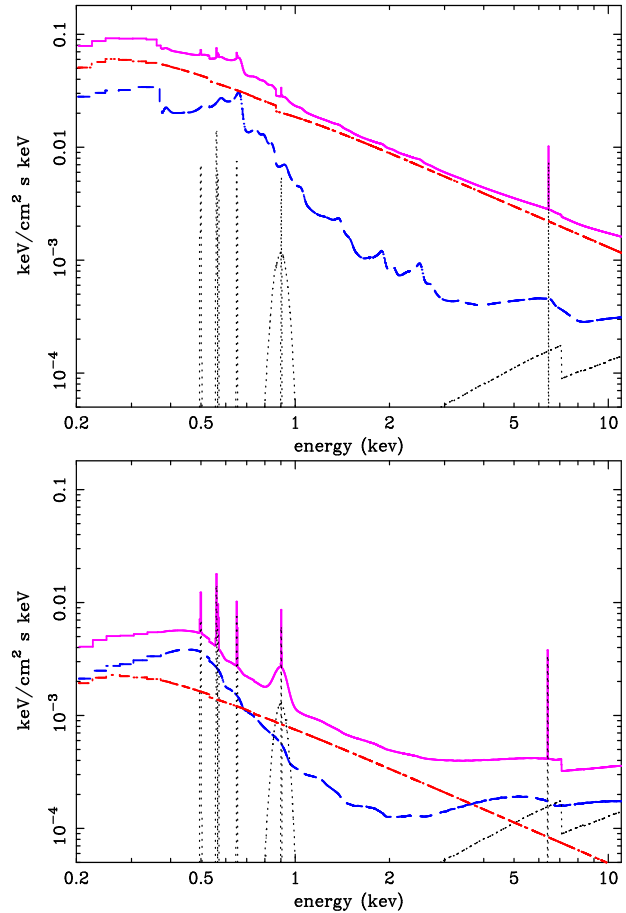

 \includegraphics[width=0.34\textwidth,height=0.46\textwidth,angle=-90]{spectrum39.cps}
 \includegraphics[width=0.34\textwidth,height=0.46\textwidth,angle=-90]{spectrum8.cps} 
 \caption{In the top panel we show the best fit model during the
   highest flux period.  The PLC is shown with a
   dash--dotted red line; the blurred ionized reflection component is
   shown with a dashed blue line; the constant reflection component
   and the emission from photionized gas is shown with dotted grey
   lines and the total emission is shown with solid magenta line.  In
   the lower panel we show the the best fit model, as before, during a
   low flux spectrum, showing the typical variability properties in
   the two--component model.}
\label{Refl_emod}
\end{figure}

The second characteristic of the model is that the soft excess is not
due to a thermal component anymore (see Fig.\ref{Refl_emod}). It is in
fact the result of including ionized reflection from the disc, which
naturally produces a soft excess. The ``constant temperature'' problem
can then be naturally solved because of the very non--thermal nature
of that component. The same kind of model successfully reproduces the
soft excesses in other sources as well (e.g. Fabian et al 2004; 2005).

The ionization parameter of the reflection spectrum is not very well
constrained by the data and most spectra are consistent with $\xi$
between 50 and 300~erg~cm~s$^{-1}$ with only few exceptions below and
above (and no clear trend with flux). The reason for the non--accurate
measurement of $\xi$ is probably twofold. Firstly, the spectra lack
sharp emission lines which would help in constraining the ionization
state (if present, the emission lines are broadened and skewed because
emitted from the inner disc). Secondly, we are using a reflection
model with a single uniform ionization parameter over all the disc
surface from the innermost to the outermost regions. This is clearly a
rough approximation for a real case in which the ionization parameter
has a structured radial profile. Such approximation is likely to play
a role in the non--accurate measurement of $\xi$.

\begin{figure}
\includegraphics[width=0.34\textwidth,height=0.46\textwidth,angle=-90]{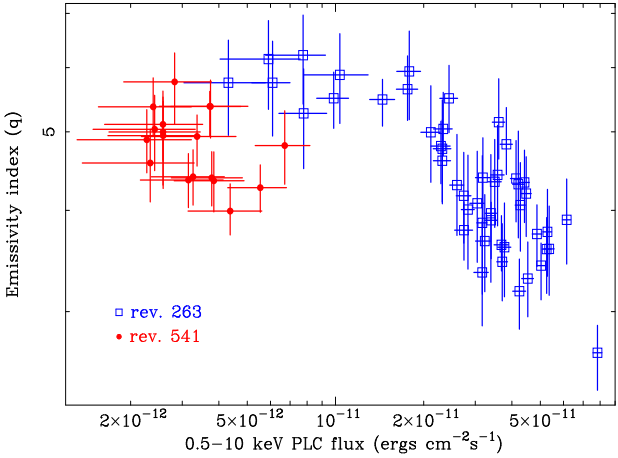}
\caption{The emissivity index of the relativistic blurring model is
  plotted versus the PLC flux in the 0.5--10~keV band.}
\label{blur}
\end{figure}

On the other hand, the disc reflection emissivity index of the
relativistic blurring model is better constrained. In Fig.~\ref{blur}
we show the emissivity index $q$ as a function of the PLC flux. Some
trend can be seen in the figure with low flux states generally
corresponding to steeper emissivity profiles (with $q$ of the order of
5) than high flux ones. As mentioned above, this result is in line
with the predictions of the light bending model (Miniutti \& Fabian 2004).

\begin{figure}
 \includegraphics[width=0.34\textwidth,height=0.46\textwidth,angle=-90]{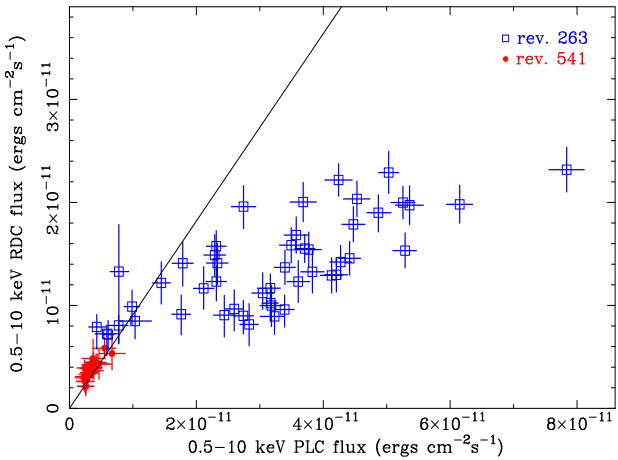} 
\caption{RDC vs. PLC 0.5--10 keV fluxes. The data clearly rule out a
  perfect correlation between the two components (solid line). The RDC
is well correlated with the PLC at low fluxes only and varies with
smaller amplitude in normal/high flux states.}
\label{Gflux}
\end{figure}

The other main prediction of the light bending model is the RDC versus
PLC relation. As mentioned, the model predicts a correlation at low
flux levels and an almost constant RDC at higher flux levels, despite
large variation in the PLC. In Fig.~\ref{Gflux} we show the
0.5--10~keV flux of the RDC versus the PLC flux in the same band. The
RDC is well correlated with the PLC at low flux levels (the solid line
represent perfect correlation between the two components). However, as
the PLC flux increases, the correlation clearly breaks down and the
RDC is much less variable (about a factor 2.5) than the PLC (about a
factor 7) in very good agreement with the qualitative predictions of
the model. The main difference between the predicted and observed
behaviour is that some residual correlation is seen at normal/high
flux (while the model predicts an almost constant RDC). Such residual
correlation can be easily interpreted by intrinsic luminosity
variations of the PLC source which are not accounted for in the model.

We finally remark that, if the light bending interpretation is
correct, the RDC versus PLC behaviour requires the primary source of
PLC to be centrally concentrated above the accretion disc within 15
gravitational radii at most from the central black hole (see Miniutti
\& Fabian 2004). In particular, the correlation between the two
component is predicted to occur only if the primary source of the PLC
is within $\sim$~5 gravitational radii from the hole. Therefore, one
consequence of this interpretation is that low flux states are
characterized by an extremely compact region of primary emission, well
inside the relativistic region around the black hole. 

\section{Reproducing the RMS spectra}

We tested both spectral models further by comparing the RMS spectra
with synthetic ones, obtained from the best--fitting models discussed
above. Obviously, the better the fits to the 2~ks spectra are,
  the more closely the RMS spectra will be matched, but the
comparison is not trivial because the $\chi^2$ obtained from the
time--resolved spectroscopy is a global quantity in the energy space
(in fact it depend over all the energies considered) while the RMS is
a local quantity (it does not depend on what is the model at
other energies). It is possible, for example, that the spectral model
slightly fails always in the same energy band. If this is the case, a
comparison between observed and simulated RMS spectra is useful.

\begin{figure}
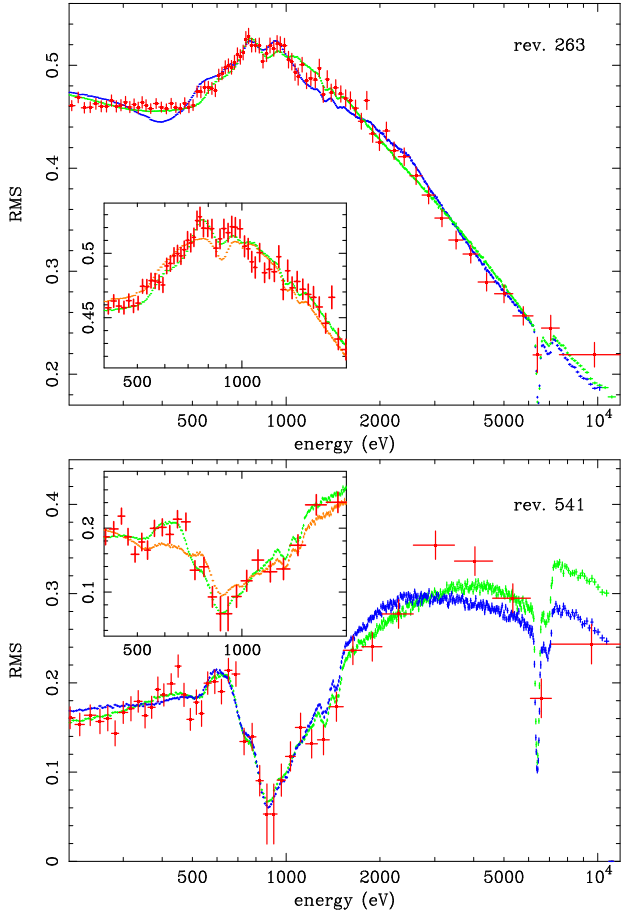

 \includegraphics[width=0.34\textwidth,height=0.46\textwidth,angle=-90]{RMS_263_fitCOL.cps}
\vspace{-3.4cm}

\hspace{0.9cm}
 \includegraphics[width=0.14\textwidth,height=0.2\textwidth,angle=-90]{RMS_263_WA.cps}

\vspace{0.9cm}

 \includegraphics[width=0.34\textwidth,height=0.46\textwidth,angle=-90]{RMS_541_fitCOL.cps}
\vspace{-5.9cm}

\hspace{0.9cm}
 \includegraphics[width=0.14\textwidth,height=0.2\textwidth,angle=-90]{RMS_541_WA.cps}

\vspace{3.3cm}

\caption{The observed RMS spectra of both observations are compared
  with synthetic ones obtained from the best--fitting standard (grey,
  green in the colour version) and two--component models (black, blue
  in colour). The inserts show the enlargement of the RMS in the soft
  energy band. The grey (green in the colour version) line shows the
  synthetic RMS obtained starting from the best--fitting parameters
  with the standard model and with varying optical depth edges,
  while the black (orange in colour) line shows the RMS expected from
  refitting the data with the imposition of a constant warm absorber.
}
\label{RMS2}
\end{figure}

In Fig.~\ref{RMS2} we show the observed and synthetic RMS spectra for
both observations. In back (blue in the colour version) we plot the
RMS spectra obtained from the two--component model best--fitting
parameters and in grey (green) that from the standard BB plus power
law model. Both models reproduce well the broadband shape of the RMS
spectra in the two observations with residuals of the order of few per
cent only.  This is possible, allowing the three parameters of
  the RDC component (ionization, emissivity index and normalization)
  to vary (see Sect. 6.3). We do not exclude the presence of further
  soft component with a possibly different physical origin, but we
  show that that is not required when variations of the parameters of
  the RDC are considered. In particular, the sharp features in the
RMS spectra have been reproduced thanks to the inclusion of the two
constant components due to distant neutral reflection and photoionized
gas emission.  This reinforces our previous qualitatively
interpretation indicating that the emission line intensities and their
constancy (not only during each observation but also between the two
one year apart) are needed in order to reproduce the RMS spectra.

The variability spectrum during the rev. 263 is less affected by the two 
constant components and carries more informations about the nuclear 
variability. 
Its shape can be explained equivalently well by the two models investigated. 
In the ``standard scenario'' the gradual drop of variability with
energy  is mainly due 
to the steepening of the power law slope  with flux, while the lower 
variability at low energy is due to a less variable BB component. 
In the two--component model scenario it is expected that the source has 
lower variability both a high and low energy, because these are 
the regions where the almost constant relativistic reflection 
component is more important.

In order to estimate the contribution, to the overall spectral
variability, due to the variations of the warm absorber, we have
simulated the RMS spectra both with constant and varying optical depth
edges.  The inserts of Figure \ref{RMS2} show the enlargement of the
RMS in the soft energy band, where the effect of the warm absorber are
stronger. The grey (green in the colour version) line shows the
synthetic RMS obtained starting from the best--fitting parameters with
the ``standard model'' and with varying optical depth edges, while the
black (orange in the colour version) line shows the RMS expected from
refitting the data with the imposition of a constant warm absorber.
We show here only the results obtained considering the standard model,
but similar results are obtained with the two component model.
The inserts of Figure \ref{RMS2} show that, when the constancy of the
optical depth is imposed, some deviations in the reproduction of the
RMS are present.  The intensity of these deviations are of the order
of few percent suggesting that a variation of the warm absorber is
possible but the bulk of the spectral variability is due to the
interplay of the back body and power law components (power law and
reflection, in the two component model) if the continuum is
left completely free to vary.

\section{Discussion}

The analysis we presented above shows that RMS spectra, flux-flux
plots and individual 2~ks spectra can be explained either in terms of
a standard back body plus power law model or of a two--component
model. In both cases, the presence of constant emission from
photoionized gas and from a distant reflector are required.

The standard model comprises a blackbody component which provides the
soft excess and a power law with variable slope (and normalization).
The blackbody temperature is surprisingly constant given that its
luminosity spans one order of magnitude and does not seem to follow
the blackbody law ($L\propto T^4$). Moreover the average temperature
we measure (about 110-120~eV) is remarkably similar to the average
temperature measured in a sample of 26 radio--quiet quasars with
typical black hole masses three orders of magnitude larger than in
NGC~4051. These surprising results suggest that a pure thermal
component is not appropriate to account for all the soft excess in
NGC~4051. In fact, as shown in Crummy et al. (in prep.) the soft
excess in type 1 active galactic nuclei is consistent, like here, with
being produced by ionized reflection. The power law slope in \ngc~
correlates with flux except at very low flux levels where the relation
expected from simple spectral pivoting breaks down. Moreover the
hardest photon index we measure (about 1.3--1.4) seem too hard to be
explained in terms of standard Comptonization models. It should be
stressed also that while spectral pivoting describes well the
flux--flux plot relationship between the 0.1--0.5~keV and 2--10~keV
bands (see Uttley et al 2004), it fails (or at least reduces its
accuracy) when the reference soft band is chosen to be between 1~keV
and 1.4~keV. This is somewhat surprising because the latter band is
dominated (whatever spectral model is preferred) by the power law
component, while at softer energies a contribution from the prominent
soft excess is unavoidable.

In the context of the two--component model, the soft excess is instead
accounted for by ionized reflection from the accretion disc which
simultaneously hardens the hard spectral shape as the flux drops. The
apparent constant temperature of the soft excess is then explained,
because the soft excess spectral shape is dictated by atomic rather
than thermal mechanisms. The two--component model is consistent with
the variability being dominated by a constant--slope power law which
varies in normalization only. The $\Gamma$--flux relationship is then
interpreted as spurious and due to the relative flux of the power law
and reflection components. The RDC is in fact correlated with the PLC
at low fluxes and varies with much smaller amplitude at normal/high
fluxes. Such behaviour explains not only the spurious $\Gamma$--flux
relationship, but is also consistent with the flux--flux plot
analysis. 

\section{Conclusions}

We investigated the X--ray spectral variability of the Narrow Line
Seyfert 1 galaxy \ngc~ with model independent techniques (RMS spectra
and Flux--Flux plots) and with time resolved spectral variability.  
The main features of the X--ray emission of
\ngc~are here summarized.

\begin{enumerate}

\item{} The \xmm~ data of \ngc~ show evidence for a distant neutral
  and constant reflection component contributing by about 10 per cent
  in the 4--10~keV band at mean fluxes.

\item{} During the low flux observation a photoionized gas imprints
  its presence in the soft X--ray band. This emission is consistent
  with being constant not only during, but also between the two
  observations one year apart. It is not clearly detected during the
  high flux observation because of a much stronger continuum. 

  The constancy of this component, as well as the previous one, is
  primarily shown thanks to the powerful RMS spectra. 

\item{} The nuclear emission has been interpreted both in a ``standard
  scenario'' (consisting of BB plus power law emission) and by a
  two--component (PLC plus ionized RDC from the disc). Both models
  reproduce the RMS spectra, and describe the time--resolved spectra
  in a comparable way from a statistical point of view. Our
    results are consistent with the warm absorber playing a minor role
    in the spectral variability.

\item{} The standard model results indicate that the BB emission does
  not follow the L$_{BB} \propto T^4$ relation, even if more than an
  order of magnitude in luminosity is spanned. The best fit BB
  temperature is almost constant, clustering its value between 0.1 and
  0.12~keV. The power law slope is consistent with spectral pivoting
  and steepens from 1.3--1.4 to about 2.2 with flux. The hardest
  photon indexes are so flat to require rather unusual scenarios such
  as a photon--starved corona. Moreover, the very low flux states
  exhibit an inverted $\Gamma$--flux behaviour which disagree with a
  simple pivoting interpretation. It is then possible that further
  unmodelled soft/steep component manifest itself at very low fluxes.

\item{} In the framework of the two--component model the soft excess
  is interpreted as the soft part of ionized reflection from the
  accretion disc. The constant temperature problem is then solved in
  terms of atomic rather than truly thermal processes. Moreover, the
  RDC explains the $\Gamma$--flux relationship at all flux levels in
  terms of the relative contribution of a PLC with constant slope
  $\Gamma=2.2$ and the RDC. The variability properties of the RDC with
  respect to the PLC explain also the reason why the flux--flux plot
  is curved without the need for invoking spectral pivoting. The RDC
  is well correlated with the PLC at low fluxes and varies with much
  smaller amplitude than it at normal/high fluxes, in line with the
  predictions of the light bending model in which the variability is
  mainly due to the fact that primary sources close to the black hole
  appear fainter than sources far from it because of light bending,
  which reduces the net observed flux at infinity as the source (of
  the PLC) is closer and closer to the hole. 

\end{enumerate}

We then propose the light bending model as a possible alternative
explanation for the spectral variability of NGC~4051 in which most of
the nuclear emission is emitted from the very inner regions of the
flow, only few gravitational radii from the central black hole. From a
physical point of view, the model does not require extreme parameters.
The only true requirement is that the PLC originates in some compact
region above the accretion disc and that the active region changes
location in time (or disappears and a new one appears at a different
location) but always within about 15 gravitational radii from the
center. We stress here that if, as it is generally believed for high
accretion rate sources (such as NGC~4051), the bulk of the X--ray
emission comes from the inner regions of the accretion flow, the
effects we are talking about are simply unavoidable. No radiation can
escape from the inner regions of the flow without being affected by
gravitational light bending and other relativistic effects. This is
true not only for disc emission (and reflection) but also for the
primary radiation.

\section*{Acknowledgments}

The authors would like to thank Mauro Dadina, Giorgio Palumbo,
Giuseppe Malaguti, Pierre Olivier Petrucci, Roderick Johnstone Michael
Mayer and the anonymous referee for many useful comments, suggestions
and technical help.  Based on observations obtained with XMM-Newton,
an ESA science mission with instruments and contributions directly
funded by ESA Member States and NASA. GP thanks the European
commission under the Marie Curie Early Stage Research Training
programme for support.  GM and KI thank the PPARC for support.  ACF
thanks the Royal Society for support.

\end{document}